\documentclass[prd,aps,twocolumn,preprintnumbers, showpacs, nofootinbib,superscriptaddress,notitlepage]{revtex4-1}
\usepackage{amssymb, amsthm}
\usepackage{amsmath}    
\usepackage{mathrsfs}   
\usepackage{graphicx}   
\usepackage{subcaption}
\usepackage{color}      
\usepackage{colortbl}
\usepackage{footnote}   
\usepackage{slashed}    
\usepackage{siunitx}    

\usepackage[utf8]{inputenc}
\usepackage[dvipsnames]{xcolor}
\usepackage[normalem]{ulem}

\usepackage{subfiles}

\newcommand{\nn}{\nonumber}

\newcommand{\beq}{\begin{equation}}
\newcommand{\eeq}{\end{equation}}

\begin{document}


\title{Quark Transverse Spin-Momentum Correlation of the Pion from Lattice QCD: The Boer-Mulders Function}


\author{\includegraphics[scale=0.1]{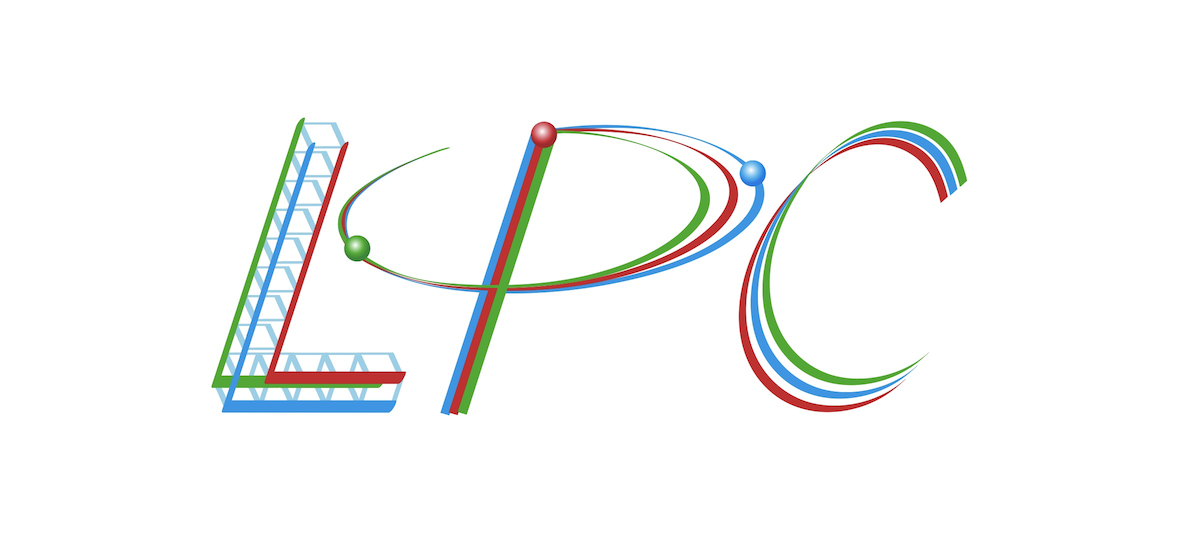}\\
Lisa Walter}
\affiliation{Institut f\"ur Theoretische Physik, Universit\"at Regensburg, D-93040 Regensburg, Germany}

\author{Jun Hua} 
\affiliation{Guangdong Provincial Key Laboratory of Nuclear Science, Institute of Quantum Matter, South China Normal University, Guangzhou 510006, China}
\affiliation{Guangdong-Hong Kong Joint Laboratory of Quantum Matter, Southern Nuclear Science Computing Center, South China Normal University, Guangzhou 510006, China}

\author{Sebastian Lahrtz}
\affiliation{Institut f\"ur Theoretische Physik, Universit\"at Regensburg, D-93040 Regensburg, Germany}

\author{Lingquan Ma}
\affiliation{Center of Advanced Quantum Studies, Department of Physics, Beijing Normal University, Beijing 100875, China}

\author{Andreas Sch\"afer}
\affiliation{Institut f\"ur Theoretische Physik, Universit\"at Regensburg, D-93040 Regensburg, Germany}

\author{Hai-Tao Shu}
\affiliation{Key Laboratory of Quark \& Lepton Physics (MOE) and Institute of Particle Physics, Central China Normal University, Wuhan 430079, China}

\author{Yushan Su}
\affiliation{Department of Physics, University of Maryland, College Park, MD 20742, USA}

\author{Peng Sun}
\affiliation{Institute of Modern Physics, Chinese Academy of Sciences, Lanzhou, Gansu Province 730000, China}

\author{Wei Wang}
\affiliation{INPAC, Key Laboratory for Particle Astrophysics and Cosmology (MOE),  Shanghai Key Laboratory for Particle Physics and Cosmology, School of Physics and Astronomy, Shanghai Jiao Tong University, Shanghai 200240, China}

\affiliation{Southern Center for Nuclear-Science Theory (SCNT), Institute of Modern Physics, Chinese Academy of Sciences, Huizhou 516000, Guangdong Province, P.R. China}

\author{Xiaonu Xiong}
\affiliation{School of Physics and Electronics, Central South University, Changsha 418003, China}

\author{Yi-Bo Yang}
\affiliation{CAS Key Laboratory of Theoretical Physics, Institute of Theoretical Physics, Chinese Academy of Sciences, Beijing 100190, China}
\affiliation{School of Fundamental Physics and Mathematical Sciences, Hangzhou Institute for Advanced Study, UCAS, Hangzhou 310024, China}
\affiliation{International Centre for Theoretical Physics Asia-Pacific, Beijing/Hangzhou, China}
\affiliation{University of Chinese Academy of Sciences, School of Physical Sciences, Beijing 100049, China}

\author{Jian-Hui Zhang}
\email{Corresponding author: zhangjianhui@cuhk.edu.cn}
\affiliation{School of Science and Engineering, The Chinese University of Hong Kong, Shenzhen 518172, China}

\author{Qi-An Zhang}
\email{Corresponding author: zhangqa@buaa.edu.cn}
\affiliation{School of Physics, Beihang University, Beijing 102206, China}

\begin{abstract}
We present the first lattice QCD calculation of the quark transverse spin-momentum correlation, i.e., the T-odd Boer-Mulders function, of the pion, using large-momentum effective theory (LaMET). The calculation is done at three lattice spacings $a=\{0.098, 0.085, 0.064\}$~fm and pion masses $\sim350$ MeV, with pion momenta up to $1.8$ GeV. The matrix elements are renormalized in a state-of-the-art scheme and extrapolated to the continuum and infinite momentum limit. We have implemented the perturbative matching up to the next-to-next-to-leading order and carried out a renormalization-group resummation. Our results provide valuable input for phenomenological analyses of the Boer-Mulders single-spin asymmetry.
\end{abstract}

\maketitle

\section{Introduction:}
Transverse-momentum-dependent parton distribution functions (TMDPDFs) encode the transverse momentum dependence of partons inside a hadron, and play a crucial role in 3D hadron tomography at the future Electron-Ion Collider (EIC) in the US~\cite{AbdulKhalek:2021gbh} and the Electron-Ion Collider in China (EicC)~\cite{Anderle:2021wcy}. At leading-twist accuracy, there exist eight quark TMDPDFs. Among them, the Sivers and the Boer-Mulders functions are time-reversal odd quantities that arise exclusively from final or initial state interactions in certain scattering processes. They characterize the unpolarized quark distribution in a transversely polarized hadron and the transversely polarized quark distribution in an unpolarized hadron, respectively. Phenomenologically, TMDPDFs can be extracted from a global fitting of TMD observables using QCD factorization theorems. While various TMDPDF fits have been performed in the literature~\cite{Bacchetta:2017gcc,Scimemi:2017etj,Bertone:2019nxa,Scimemi:2019cmh,Bacchetta:2019sam,Bacchetta:2020gko,Bury:2021sue,Echevarria:2020hpy}, most of them have focused on the unpolarized quark TMDPDF, with limited analyses on the Sivers function, leaving the T-odd Boer-Mulders function relatively underexplored. Lattice QCD has the potential to play an important complementary role in calculating TMDPDFs, particularly the T-odd Sivers and Boer-Mulders functions, from first principles.

Early lattice studies have been focused on accessing the moments of TMDPDFs~\cite{Hagler:2009mb,Musch:2010ka,Musch:2011er,Engelhardt:2015xja,Yoon:2017qzo}, including the Sivers and Boer-Mulders functions~\cite{Musch:2011er,Engelhardt:2015xja}, by studying ratios of suitable correlators. With the proposal of LaMET~\cite{Ji:2013dva,Ji:2014gla,Ji:2020ect}, we are also able to access the full TMDPDFs. In the past few years, there has been rapid progress in both theoretical developments~\cite{Ji:2014hxa,Ji:2018hvs,Ji:2019sxk,Ji:2019ewn,Ji:2020jeb,Zhang:2020dbb,Ebert:2018gzl,Ebert:2019okf,Ebert:2019tvc,Shanahan:2019zcq,Ebert:2020gxr,Vladimirov:2020ofp,Ji:2021uvr,Ebert:2022fmh,Schindler:2022eva,Zhu:2022bja,Zhao:2023ptv,Han:2024min} and lattice calculations~\cite{Shanahan:2020zxr,LatticeParton:2020uhz,Schlemmer:2021aij,Shanahan:2021tst,LatticePartonLPC:2022eev,LatticePartonCollaborationLPC:2022myp,LatticeParton:2023xdl,Avkhadiev:2023poz,Avkhadiev:2024mgd,LatticeParton:2018gjr,Han:2024yun}. On the theory side, it has been well established how to connect the quark and gluon TMDPDFs to appropriately defined quasi-light-front (quasi-LF) correlations or quasi-TMDPDFs involving staple-shaped Wilson line operators using a factorization formula~\cite{Ebert:2019okf,Ji:2019ewn,Ebert:2020gxr,Vladimirov:2020ofp,Ebert:2022fmh,Schindler:2022eva,Zhu:2022bja}. The perturbative matching has been calculated up to the next-to-next-to-leading order~\cite{delRio:2023pse,Ji:2023pba}. In particular, the factorization and perturbative matching takes a universal form for all leading-twist TMDPDFs~\cite{Ebert:2020gxr,Schindler:2022eva,Zhu:2022bja}. 
There have also been studies examining the potential higher-twist contamination in the extraction of leading-twist quark TMDPDFs~\cite{Vladimirov:2020ofp}, suggesting that the effect may be the weakest for the T-odd Boer-Mulders function.
On the lattice side, various calculations have been conducted of the soft function~\cite{LatticeParton:2020uhz}, the Collins-Soper kernel~\cite{Shanahan:2020zxr,Schlemmer:2021aij,Shanahan:2021tst,LatticePartonLPC:2022eev,Avkhadiev:2023poz,Avkhadiev:2024mgd} controlling the rapidity evolution of TMDPDFs, as well as the unpolarized quark TMDPDF~\cite{LatticePartonCollaborationLPC:2022myp} and the pion TMD wave function~\cite{LatticeParton:2023xdl}. 

In this work, we perform a systematic study of the T-odd Boer-Mulders function of the pion using LaMET. The lattice matrix elements of the Boer-Mulders quasi-LF correlator are calculated with three lattice spacings $a=\{0.098, 0.085, 0.064\}$~fm, pion masses $\sim350$ MeV, and pion momenta up to $1.8$ GeV. The matrix elements are then renormalized in a state-of-the-art scheme~\cite{Zhang:2022xuw} and Fourier transformed to longitudinal momentum space. Using the soft function and Collins-Soper kernel calculated previously~\cite{LatticePartonLPC:2023pdv}, we extract the Boer-Mulders function and extrapolate the result to the continuum and infinite momentum limit. We have implemented the perturbative matching up to the NNLO and carried out a renormalization-group resummation.

The rest of the paper is organized as follows: In Sec.~\ref{SEC:qTMD}, we give a brief overview of the theoretical framework that allows us to extract the Boer-Mulders function from the quasi-Boer-Mulders (quasi-BM) function. We then discuss in Sec.~\ref{SEC:latt} details of our lattice calculation. Sec.~\ref{SEC:numres} presents our numerical results. Finally, we conclude in Sec.~\ref{SEC:summ}.

\section{Theoretical framework}
\label{SEC:qTMD}
For spin-$0$ hadrons like the pion, there exist only two leading-twist quark TMDPDFs, rather than eight for spin-$1/2$ hadrons. One is the unpolarized TMDPDF, and the other is the Boer-Mulders function.
To access them on the lattice, we can study the following subtracted quasi-TMDPDF defined in LaMET~\cite{Ji:2014hxa,Ji:2018hvs,Ebert:2019okf,Ji:2019ewn}
\begin{align}\label{eq:quasi_TMD}
& \tilde f_\Gamma(z ,b_\perp,1/a,P^z) \\
&=\! \lim_{L \rightarrow \infty}  \frac{\langle \pi(P)| \bar \psi\big(0,\vec{0}_\perp\big)\Gamma{W}_\sqsubset(\vec z, \vec{b}_\perp;\vec L)\psi\big(z, \vec b_\perp\big) |\pi(P)\rangle}{\sqrt{Z_E(2L+z,b_\perp,1/a)}} \ , \nonumber
\end{align}
where $|\pi(P)\rangle$ denotes a pion with momentum $P=(P^0,0,0,P^z)$, and $a$ is the lattice spacing. $\Gamma=\{\gamma^t,\gamma^t\gamma^{\perp}\gamma_5\}$ defines the unpolarized and Boer-Mulders quasi-TMDPDFs $\tilde f_1, \tilde h_1^\perp$ as following,
\begin{align}
\tilde f_{\gamma^t}&=\tilde f_1,\nn\\
\tilde f_{\gamma^t\gamma^{\perp i}\gamma_5}&=i\epsilon^{ij}_\perp b_{\perp j}M_\pi \tilde h_1^\perp,
\end{align}
where $i, j$ are transverse indices. $\vec L\equiv L  n_z$, $\vec z=z n_z$ with $ n_z=(0,0,0,1)$ being a unit four-vector along the spatial $z$ direction, and $\vec{b}_\perp=(0,b_1,b_2,0)$. In our calculation, we choose $b_2=0$. $W_\sqsubset$ denotes a staple-shaped Wilson line
\begin{align}\label{eq:staplez}
{W_\sqsubset}(\vec z, \vec{b}_\perp;L)&={W_z^\dagger\Big(\vec L+\vec z; -L-z\Big)}
\\
&\times W_{\perp}\Big(\vec L +\vec z+\vec b_\perp; -b_\perp\Big)
W_{z}\Big(\vec z+\vec b_\perp;L\Big), \nonumber\\
W_{i}(\vec\eta; L)&= {\cal P}{\rm exp}\Big[-ig\int_{0}^{L} dt\, {n}_i\cdot A(\eta^\mu+t n_i^\mu)\Big].\nn
\end{align}
$\sqrt{Z_E(2L+z,b_\perp,1/a)}$ is the square root of the vacuum expectation value of a flat rectangular Euclidean Wilson-loop along the $n_z$ direction with length $2L+z$ and width $b_\perp$:
\begin{align}\label{eq:Z_E}
Z_E(2L+z,b_\perp,1/a)&=\frac{1}{N_c}{\rm Tr}\langle 0|W_\perp^\dagger(-\vec L+\vec b_\perp;-b_\perp)\\
&\times W_z^\dagger(\vec L+\vec z+\vec b_\perp;-2L-z)\nonumber\\
&{\times W_{\perp}(\vec L+\vec z; b_\perp) W_z(-\vec L;2L+z)}|0\rangle \,\nn .
\end{align}
It serves to remove the linear and cusp divergences associated with the Wilson lines in the quark quasi-TMDPDF operator in the numerator on the r.h.s. of Eq.~(\ref{eq:quasi_TMD}). The subtracted quasi-TMDPDF then has a smooth $L\to\infty$ limit. 
However, it still contains logarithmic divergences from the endpoints of the operator, which can be removed in the short-distance ratio scheme~\cite{Zhang:2022xuw} by further dividing by the zero-momentum hadron matrix element of the same operator at small $z$ and $b_\perp$. Then the renormalized quasi-TMDPDF can be converted to the $\overline{\rm MS}$ scheme. Here we combine the renormalization and conversion factors into a single factor $Z_O$ as
\beq
Z_O(1/a,\mu)=\frac{\tilde f_\Gamma(z_0,b_{\perp,0},1/a,0)}{\tilde f_\Gamma^{\overline{\rm MS}}(z_0,b_{\perp,0},\mu,0)},
\eeq
where $\tilde f_\Gamma(z_0,b_{\perp,0},1/a,0)$ is the zero momentum matrix element calculated on the lattice, following the definition in Eq.~(\ref{eq:quasi_TMD}). $\tilde f_\Gamma^{\overline{\rm MS}}(z_0,b_{\perp,0},\mu,0)$ is the conversion factor to the $\overline{\rm MS}$ scheme with $\mu$ being the renormalization scale. In this ratio, the intrinsic dependence on $z_0$ and $b_{\perp,0}$ is expected to cancel, leaving the remaining dependence on $a$ and $\mu$ in $Z_O$. However, in practical calculations the cancellation on $z_0$ and $b_{\perp,0}$ dependence is never complete due to missing higher-order contributions. For $z_0 \ll 1/\Lambda_{\rm QCD}$ and $a < b_{\perp,0} \ll 1/\Lambda_{\rm QCD}$, $\tilde f_\Gamma^{\overline{\rm MS}}(z_0,b_{\perp,0},\mu,0)$ can be calculated in the continuum perturbation theory. It is expected to be the same for the Boer-Mulders function and for the unpolarized quark quasi-TMDPDF~\cite{Ebert:2020gxr}, and the result up to one-loop accuracy is~\cite{Zhang:2022xuw}
\begin{align}
\tilde f_\Gamma^{\overline{\rm MS}}(z_0,b_{\perp,0},\mu,0) = & 1 + \frac{\alpha_s C_F}{2\pi} \left[\frac{3}{2}\ln\left(\frac{(z_0^2+b_{\perp,0}^2)\mu^2 e^{2\gamma_E}}{4}\right) \right. \nonumber\\
&\left.-\frac{2z_0}{b_{\perp, 0}}\arctan\left(\frac{z_0}{b_{\perp, 0}}\right)+\frac{1}{2}\right] \ ,
\end{align}
where $\alpha_s=g^2/(4\pi)$ is the QCD running coupling.
To reduce the dependence on $z_0$ or $b_{\perp, 0}$, one can perform a renormalization group (RG) resummation, 
\begin{align}
\tilde f_\Gamma^{\overline{\rm MS}}(z_0,b_{\perp,0},\mu,0) 
= &\tilde f_\Gamma^{\overline{\rm MS}}(z_0,b_{\perp,0},\mu_0,0) \nonumber\\
&\times \exp\left[ \int^{\mu}_{\mu_0} \frac{d\mu'}{\mu'} 2 \gamma_{F}(\alpha_s(\mu')) \right] \ ,
\end{align}
which evaluates the fixed-order perturbative series at the physical scale $\mu_0 = r \cdot 2 e^{-\gamma_E} / \sqrt{b_{\perp,0}^2 + z_0^2}$, and evolves it to the renormalization scale $\mu$. One can vary $r=0.8\sim 1.2$ as an estimate of perturbative uncertainties. $\gamma_{F}$ is the heavy-light quark UV anomalous dimension~\cite{Ji:1991pr, Chetyrkin:2003vi, Braun:2020ymy, Grozin:2023dlk}. 

The fully renormalized quasi-TMDPDF is then given by
\beq\label{eq:renqTMD}
\tilde f_\Gamma^{\overline{\rm MS}}(z,b_\perp,\mu, P^z)=Z_O^{-1}\tilde f_\Gamma(z ,b_\perp,1/a,P^z),
\eeq
where we have assumed a continuum limit on the l.h.s.
From Eq.~(\ref{eq:renqTMD}), the momentum space density is given by the following Fourier transform
\beq \label{eq:TMD-mom}
 \tilde f_\Gamma(x, b_\perp,\mu,\zeta_z) = \int\frac{d\lambda}{2\pi}e^{ix\lambda}\tilde f_\Gamma^{\overline{\rm MS}}(\lambda/P^z,b_\perp,\mu,P^z),
\eeq
with $\lambda=z P^z$ being the quasi-LF distance and $\zeta_z=(2xP^z)^2$ the Collins-Soper scale. The dependence of $\tilde f_\Gamma$ on $\mu$ is controlled by the RG equation~\cite{Collins:1981uk,Ji:2004wu}
\beq\label{eq:RG_TMD}
\mu^2\frac{d}{d\mu^2}\ln \tilde f_\Gamma(x, b_\perp,\mu,\zeta_z)=\gamma_F(\alpha_s(\mu)).
\eeq
The dependence on $\zeta_z$ characterizes how the quasi-TMDPDF changes with momentum or rapidity, and the evolution is controlled by the Collins-Soper equation~\cite{Collins:1981uk,Ji:2014hxa}
\beq\label{eq:CS_TMD}
P^z\frac{d}{dP^z}\ln \tilde f_\Gamma(x, b_\perp, \mu, \zeta_z)=K(b_\perp,\mu)+G(\zeta_z,\mu),
\eeq
where $K(b_\perp,\mu)$ is the Collins-Soper kernel that is independent of the rapidity regulator, while $G(\zeta_z,\mu)$ is a perturbative term existing only in the off-light-cone regularization scheme, its explicit expression at one-loop can be found in Ref.~\cite{Ji:2019ewn}. 

The renormalized quasi-TMDPDF $\tilde{f}_\Gamma(x,b_{\perp},\mu,\zeta_z)$ is then related to the standard TMDPDF $f_\Gamma(x,b_{\perp},\mu,\zeta)$ by the factorization \cite{Ji:2019ewn,Ebert:2022fmh}:
\begin{align}\label{eq:matching_formula}
\tilde{f}_\Gamma(x,b_{\perp}&,\mu,\zeta_z) \sqrt{S_I(b_{\perp},\mu)} = H\left(\frac{\zeta_z}{\mu^2}\right) e^{\frac{1}{2} \ln{\left(\frac{\zeta_z}{\zeta}\right)} K(b_{\perp},\mu)}\nonumber\\
&\times f_\Gamma(x,b_{\perp},\mu,\zeta) + \mathcal{O}\left(\frac{\Lambda^2_{\text{QCD}}}{\zeta_z},\frac{M^2}{(P^z)^2},\frac{1}{b^2_{\perp}\zeta_z}\right),
\end{align}
where $S_I(b_{\perp},\mu)$ is the intrinsic or reduced soft function that can be calculated on the lattice using a light-meson form factor~\cite{Ji:2019sxk}, for example, in \cite{LatticePartonLPC:2023pdv} for the CLS ensemble X650. The rapidity scale is denoted by $\zeta=(2xP^+)^2$ and the pion mass by $M$. The power corrections $\mathcal{O}\left(\frac{\Lambda^2_{\text{QCD}}}{\zeta_z},\frac{M^2}{(P^z)^2},\frac{1}{b^2_{\perp}\zeta_z}\right)$ are suppressed for large values of the pion momentum $P^z$ and Collins-Soper scale $\zeta_z$, but will get important in the endpoint regions where LaMET becomes unreliable.

The matching kernel is often written as $H=e^h$ and has been calculated perturbatively up to the next-next-to-leading order (NNLO), where the next-to-leading order (NLO) result reads \cite{Ji:2018hvs,Ebert:2019okf,Ji:2020ect}
\begin{equation}
h^{\left( 1 \right)} \left( \frac{\zeta_z}{\mu^2} \right) = \frac{\alpha_s C_F}{2 \pi} \left( -2 + \frac{\pi^2}{12} + \ln \frac{\zeta_z}{\mu^2} - \frac{1}{2} \ln^2 \frac{\zeta_z}{\mu^2} \right),
\label{eq:TMDPDF_matching_kernel_NLO}
\end{equation}
while the NNLO result reads \cite{delRio:2023pse,Ji:2023pba}
\begin{align}\label{eq:TMDPDF_matching_kernel_NNLO}
h^{\left( 2 \right)} &\left( \frac{\zeta_z}{\mu^2} \right) = 
\alpha_s^2 \Big[ c_2 - \frac{1}{2} \left( \gamma_C^{\left( 2 \right)} - \beta_0 c_1 \right) \ln \frac{\zeta_z}{\mu^2}\nonumber\\
&- \frac{1}{4} \left( \Gamma_{\text{cusp}}^{\left( 2 \right)} - \frac{\beta_0 C_F}{2 \pi}\right) \ln^2 \frac{\zeta_z}{\mu^2} - \frac{\beta_0 C_F}{24 \pi} \ln^3 \frac{\zeta_z}{\mu^2} \Big].
\end{align}
The constants $c_2 = 0.0725 C_F^2 - 0.0840 C_F C_A + 0.1453 C_F n_f/2$ and $c_1 = \frac{C_F}{2\pi} \left(-2+\frac{\pi^2}{12}\right)$. $\beta_0=-\frac{1}{2 \pi}\left(\frac{11}{3} C_A-\frac{2}{3} n_f\right)$ is the leading order QCD beta function. $\Gamma_{\text {cusp }}^{(2)}=C_F C_A\left(-\frac{1}{12}+\frac{67}{36 \pi^2}\right)-\frac{5 C_F n_f}{18 \pi^2}$ denotes the two-loop cusp anomalous dimension~\cite{Korchemskaya:1992je}. The single log anomalous dimension at two-loop is $\gamma_C^{(2)}=a_1 C_F C_A + a_2 C_F^2 + a_3 C_F n_f$ with coefficients $a_1 = 44 \zeta_3 - \frac{11\pi^2}{3} - \frac{1108}{27}$, $a_2 = -48 \zeta_3 + \frac{28 \pi^2}{3} - 8$ and $a_3 = \frac{2 \pi^2}{3} + \frac{160}{27}$~\cite{Ji:2019ewn,Ji:2020ect}. 

The hard kernel $H$ satisfies the following RG equation~\cite{Ji:2019ewn,Ji:2023pba},
\begin{align}
\mu \frac{d}{d \mu} \ln H\left(\frac{\zeta_z}{\mu^2}\right)=\Gamma_{\text {cusp }} \ln \frac{\zeta_z}{\mu^2}+\gamma_C \ ,
\end{align}
which indicates an all-order mathematical structure of double logarithms $\sim \alpha_{s}^n \ln^{2n}\left(\zeta_z/\mu^2\right)$, and can be numerically large if $\zeta_z$ differs from $\mu^2$ by an appreciable amount, such as in the small $x$ region. Therefore, to perform the perturbative matching in a reliable and controllable manner, especially for the small $x$ region, one has to resum the double logarithms to all orders in perturbation series.  

The standard resummation approach starts from evaluating the fixed order perturbation series at the physical scale $r^2\zeta_z$ with $r$ being an empirical parameter, and then evolves it to the factorization scale $\mu^2$ using the RG equation~\cite{Ji:2019ewn}:
\begin{align}
& H\left(\alpha_s(\mu), \frac{\zeta_z}{\mu^2}\right)=H\left(\alpha_s\left(r \sqrt{\zeta_z}\right), \frac{1}{r^2}\right) \nonumber\\
& \times \exp \left\{\int_{r\sqrt{\zeta_z}}^\mu \frac{d \mu^{\prime}}{\mu^{\prime}}\left[\Gamma_{\text {cusp }}\left(\alpha_s\left(\mu^{\prime}\right)\right) \ln \frac{\zeta_z}{\mu^{\prime 2}}+\gamma_C\left(\alpha_s\left(\mu^{\prime}\right)\right)\right]\right\} \ .
\end{align}
The physical scale is only defined to the order of magnitude, rather than as a precise value. We can vary the parameter $r=0.8\sim1.2$ to estimate the uncertainties arising from the physical scale choice. If one knows the hard kernel to all orders, the variation of $r$ does not influence the result. Since we only know the hard kernel up to limited orders, varying $r$ could lead to notable effects, which can be viewed as estimated uncertainties due to unknown higher-order perturbative corrections. 

The resummation accuracy depends on the order of the fixed-order results and the anomalous dimensions, which is shown in Table~\ref{appschS}. Calculations of the cusp anomalous dimension $\Gamma_{\text{cusp}}$ for the quark case are up to four-loop~\cite{Korchemskaya:1992je,Moch:2004pa,Henn:2019swt,vonManteuffel:2020vjv,Grozin:2022umo}. The single log anomalous dimension $\gamma_C$ up to three-loop is presented in Ref.~\cite{Ji:2024hit}, which is obtained based on the universality of anomalous dimensions discussed in Ref.~\cite{Ji:2023pba} and the perturbative results in Ref.~\cite{Becher:2006mr,Braun:2020ymy,Bruser:2019yjk,Moult:2022xzt}. The current results in the literature allow for NNLO+RGR accuracy. 
\begin{table}[htbp]
\begin{equation}
\begin{array}{|c|c|c|c|c|}
\hline \text {RG scheme} & \text {Accuracy}\sim \alpha_s^n L_{z}^k & \Gamma_{\text {cusp}} & \gamma_C & \text{fixed order} \\
\hline-  & k = 2 n \,  & \text {1-loop} & \text {tree} & \text {tree} \\
\hline \mathrm{LO+RGR}  & 2n-1 \leq k \leq 2 n  & \text {2-loop} & \text {1-loop} & \text {tree} \\
\hline \mathrm{NLO+RGR}  & 2n-3 \leq k \leq 2 n  &  \text {3-loop} & \text {2-loop} & \text {1-loop} \\
\hline \mathrm{NNLO+RGR}  & 2n-5 \leq k \leq 2 n  & \text {4-loop} &  \text {3-loop} & \text {2-loop} \\
\hline
\end{array} \nonumber
\end{equation}
\caption{Different resummation accuracies for the hard kernel $H$. The log term is $L_z = \ln\left(\zeta_z/\mu^2\right)$. }
\label{appschS}
\end{table}


\section{Lattice calculations}

\label{SEC:latt}

\subsection{Lattice setup}
Our calculation uses three different lattice ensembles generated by the CLS collaboration~\cite{Bruno:2014jqa} with lattice spacing $a=\{0.098,0.085,0.064\}~\mathrm{fm}$ and pion masses $\sim 350$ MeV. For each ensemble, we calculate several source-sink separations with eight sources per configuration. Four individual measurements are used for each configuration for X650, while for H102 and N203, eight measurements are performed. Details of the lattice setup and parameters are collected in Table~\ref{Tab:setup}. The pion momenta used in the calculation are $P^z=\{0,0.53,0.79,1.05,1.32,1.58,1.84\}\si{\giga\electronvolt}$ for X650, $P^z=\{0,0.91,1.37,1.82\}\si{\giga\electronvolt}$ for H102 and $P^z=\{0,0.81,1.21,1.61\}\si{\giga\electronvolt}$ for N203. The number of configurations used for $P^z=\{0,0.81\}\si{\giga\electronvolt}$ is reduced to 500 to save computing resources, while still a sufficient signal can still be seen with the smaller subset of configurations.

\begin{table}
\centering
\begin{tabular}{cclccccccc}
\hline
\hline
Ensemble ~&$a$(fm) ~& \ \!$L^3\times T$  ~& $m_\pi$(MeV) ~& $m_\pi L$  & $N_\mathrm{conf.}$\\
\hline
X650  ~& 0.098  ~& $48^3\times 48$  ~& 338   ~&8.1    &1892                                \\ 
H102  ~& 0.085  ~& $32^3\times 96$  ~& 354   ~&4.9    &1008                 \\ 
N203  ~& 0.064  ~& $48^3\times 128$  ~& 348  ~&5.4   &500/1543      \\ 
\hline
\end{tabular}
 \caption{The simulation setup, including lattice spacing $a$, lattice size $L^3\times T$, pion masses~\cite{Bali:2023sdi} and the number $N_{\text{conf.}}$ of configurations. } 
 \label{Tab:setup}
\end{table}

\begin{figure}[thbp]
\includegraphics[width=.45\textwidth]{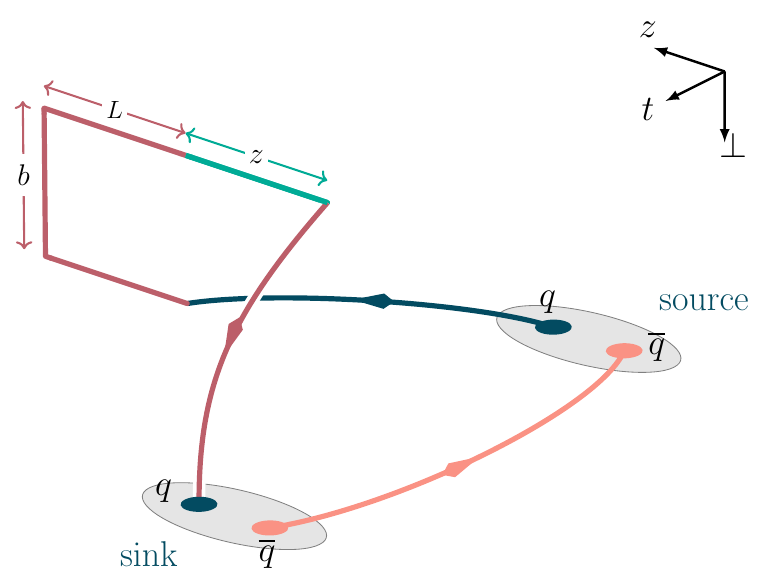}
\caption{Illustration of the lattice setup for the three-point function. The sequential source method \cite{Martinelli:1988rr} is used.} 
\label{fig:setup_threepoint_function}
\end{figure}

A schematic illustration of the lattice setup for calculating the three-point function needed to obtain the matrix element in the numerator of the quasi-TMDPDF defined in Eq. \eqref{eq:quasi_TMD} is shown in Fig. \ref{fig:setup_threepoint_function}. The sequential source method \cite{Martinelli:1988rr} is used. 

The source-sink separations and maximum $z$ and $b_{\perp}$ as well as the staple link length $L$ used in the calculation of the three-point correlation function are listed in Table \ref{tab:tseP^zmax_bmax}. 

\begin{table}
    \centering
    \begin{tabular}{ccccc}
        \hline
        \hline
        Ensemble & $t_{\text{sep}}/a$ & $z_{\text{max}}/a$ & $b_{\perp, \text{max}}/a$ & L/a \\
        X650 & \{6,7,8,9,10\} & 18 & 7 & \{8,10\} \\
        H102 & \{7,8,9,10,11,12\} & 22 & 8 & \{8,10\} \\
        N203 & \{9,11,13,15,17\} & 28 & 9 & \{10,12\} \\
        \hline
        \hline
    \end{tabular}
    \caption{Source-sink separations $t_{\text{sep}}/a$, maximum longitudinal and transverse separations $z_{\text{max}}/a$ and $b_{\perp, \text{max}}/a$ of the quark fields in the quasi-TMDPDF defined in eq. \eqref{eq:quasi_TMD} and staple link length $L/a$ used in the analysis.}
    \label{tab:tseP^zmax_bmax}
\end{table}

\subsection{Dispersion relation}

\begin{figure}[thbp]
\includegraphics[width=.45\textwidth]{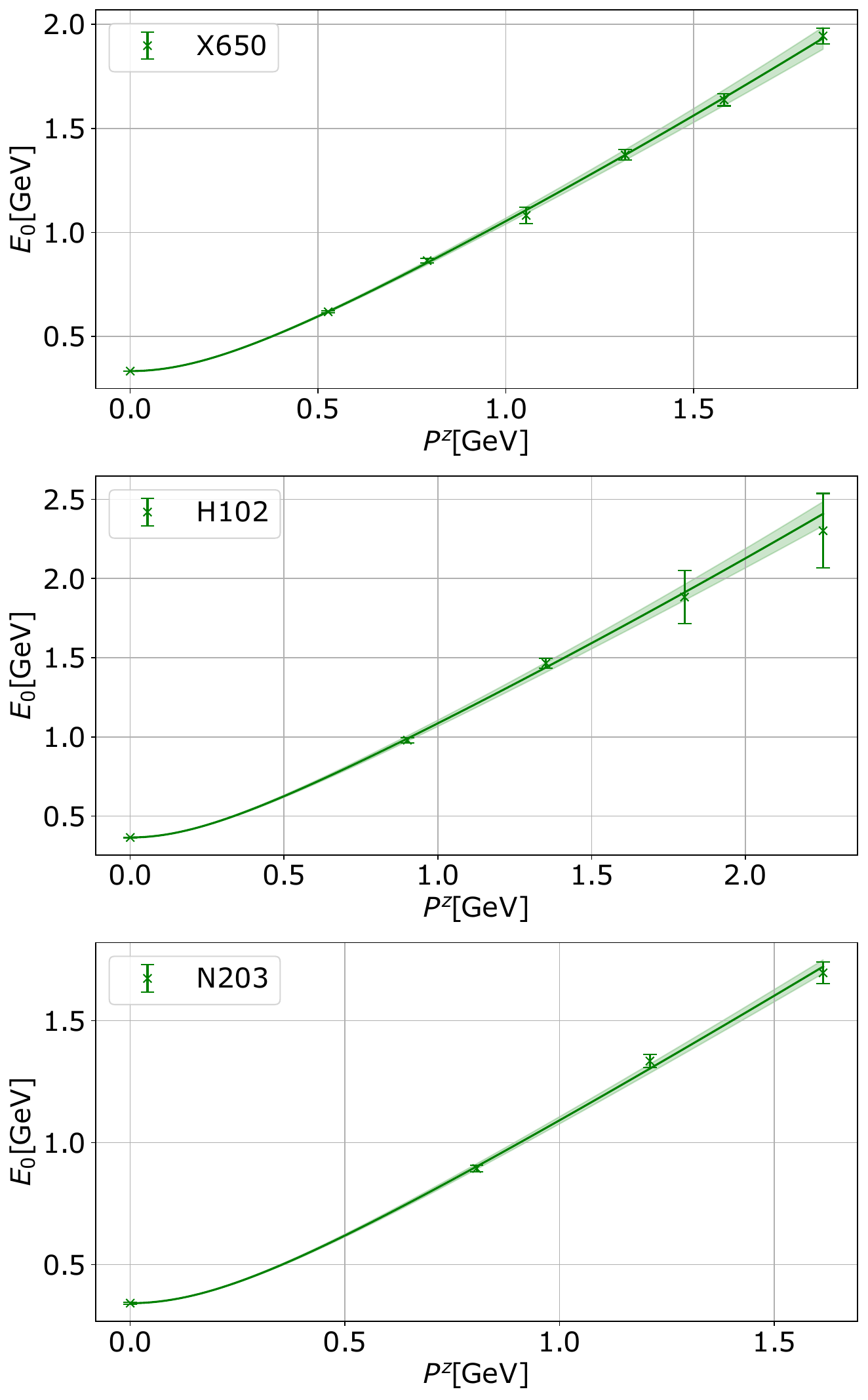}
\caption{Dispersion relations for the pion on three different ensembles.} 
\label{fig:Disp-rel}
\end{figure}

Using the two-state parametrization 

\begin{align}\label{eq:twopoint_fit}
C^{\text{2pt}}(P^z, t_{\text{sep}}) = c_4 e^{-E_0 t_{\text{sep}}}(1 + c_5 e^{-\Delta E t_{\text{sep}}}),
\end{align}
where $t_{\rm {sep}}$ is the source-sink separation, the ground state energies $E_0$ can be extracted from fitting the two-point function. In the equation above, the dependence of the fit parameters $c_4$, $c_5$, $E_0$ and $\Delta E$ on $P^z$ has been omitted for simplicity. From the ground-state energies at different momenta for all ensembles, the pion dispersion relation can be obtained for different lattice spacings. We parametrize this relation with

\begin{align}\label{eq:dispersion_relation}
E_0(P^z) = \sqrt{m^2 + c_1 (P^z)^2 + c_2 (P^z)^4 a^2},
\end{align}
where discretization errors are taken into account by the term quadratic in $a$. In Fig.~\ref{fig:Disp-rel}, we plot the dispersion relations for the three ensembles. The fit results in $c_1=\num{0.972 \pm 0.018}$ and $c_2 = \num{0.111 \pm 0.050}$ for X650 and $c_1 = \num{1.028 \pm 0.036}$ and $c_2 = \num{0.093 \pm 0.038}$ for H102. For N203, the dispersion relation fit gives $c_1 = \num{1.063 \pm 0.028}$ and $c_2 = \num{0.106 \pm 0.031}$. The results indicate good agreement with the continuum dispersion relation $E_0(P^z) = \sqrt{m^2+(P^z)^2}$.

To estimate the uncertainties, binning with bin size 5 and bootstrap resampling with 800 samples is used. The bootstrap samples and correlations are kept consistent during the entire analysis.

\subsection{Joint fit of two- and three-point functions}
To extract the ground state matrix element, we decompose the two-point correlator $C^{2\text{pt}}(P^z,t_\text{sep})$ and three-point correlator $C_\Gamma^{3\text{pt}} (P^z, t, t_\text{sep})$ (with $\Gamma=\gamma^t\gamma^\perp \gamma_5$) as in~\cite{Bhattacharya:2013ehc}, 
%
\begin{align}\label{eq:twostate}
C^\text{2pt}(P^z,t_\text{sep}) &=|{\cal A}_0|^2 e^{-E_0t_\text{sep}}+|{\cal A}_1|^2 e^{-E_1t_\text{sep}}+\cdots\,,\nn\\
C^\text{3pt}_{\Gamma}(P^z,t,t_\text{sep}) &=
   |{\cal A}_0|^2 \langle 0 | {O}_\Gamma | 0 \rangle  e^{-E_0t_\text{sep}} \nonumber\\
   &+|{\cal A}_1|^2 \langle 1 | {O}_\Gamma | 1 \rangle  e^{-E_1t_\text{sep}} \nonumber\\
   &+{\cal A}_1{\cal A}_0^* \langle 1 | {O}_\Gamma | 0 \rangle  e^{-E_1 (t_\text{sep}-t)} e^{-E_0 t} \nonumber\\
   &+{\cal A}_0{\cal A}_1^* \langle 0 | {O}_\Gamma | 1 \rangle  e^{-E_0 (t_\text{sep}-t)} e^{-E_1 t} + \cdots \,,
\end{align}
where $\left\langle 0 \left| \mathcal{O}_\Gamma \right|  0\right\rangle=\tilde{h}_1^{\perp,0}(z,b_{\perp},1/a,P^z)$ is the ground state matrix element corresponding to the numerator of the r.h.s. of Eq.~(\ref{eq:quasi_TMD}), $t$ denotes the insertion time of $O_\Gamma$. The ellipses denote the contribution from higher excited states of the pion which decay faster than the ground state and first-excited state. We extract the bare ground state matrix element by performing a two-state combined fit with $C^{2\text{pt}}(P^z,t_\text{sep})$ and the ratio  $R_\Gamma(P^z,t_\text{sep},t)={C_\Gamma^{3\text{pt}}(P^z,t_\text{sep},t)}/{C^{2\text{pt}}(P^z,t_\text{sep})}$. 

For the combined fit, the two-point function is parametrized by Eq. \eqref{eq:twopoint_fit} and the ratio with
\begin{align}\label{eq:fit_function_ratio}
    R_\Gamma(t_\text{sep},t) = \frac{\tilde{h}_1^{\perp,0} + c_1 [e^{-\Delta E t} + e^{-\Delta E (t_{\text{sep}} - t)}] + c_3 e^{-\Delta E t_{\text{sep}}}}{1 + c_5 e^{-\Delta E t_{\text{sep}}}}.
\end{align}
The bare ground state matrix element $\tilde{h}_1^{\perp,0}(z,b_{\perp},1/a,P^z)$ is extracted by performing a correlated joint fit at fixed values of $z$, $b_{\perp}$ and $P^z$ for each ensemble. The insertion times $t=0$ and $t=t_{\text{sep}}$ are excluded for the three-point functions in the fitting process for X650 and H102, while for N203 the points with $t=1$ and $t=t_{\text{sep}}-1$ are also excluded.

\begin{figure*}[thbp]
\vspace{-1em}
\begin{subfigure}{0.32\textwidth}
\includegraphics[width=\textwidth]{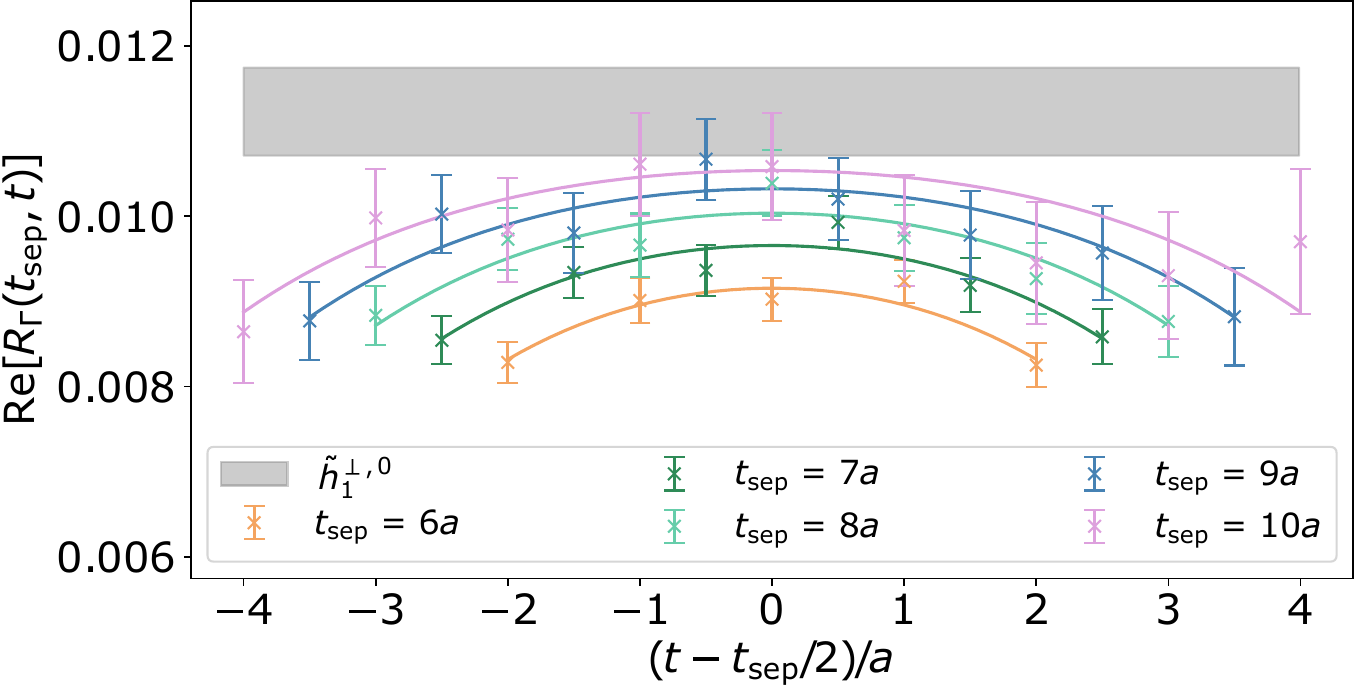}
\caption{X650, $P^z=0.79$~GeV, $z=5a$, $b_{\perp}=3a$}
\label{ratio_fit_X650-1}
\end{subfigure}
\hfill
\begin{subfigure}{0.32\textwidth}
\includegraphics[width=\textwidth]{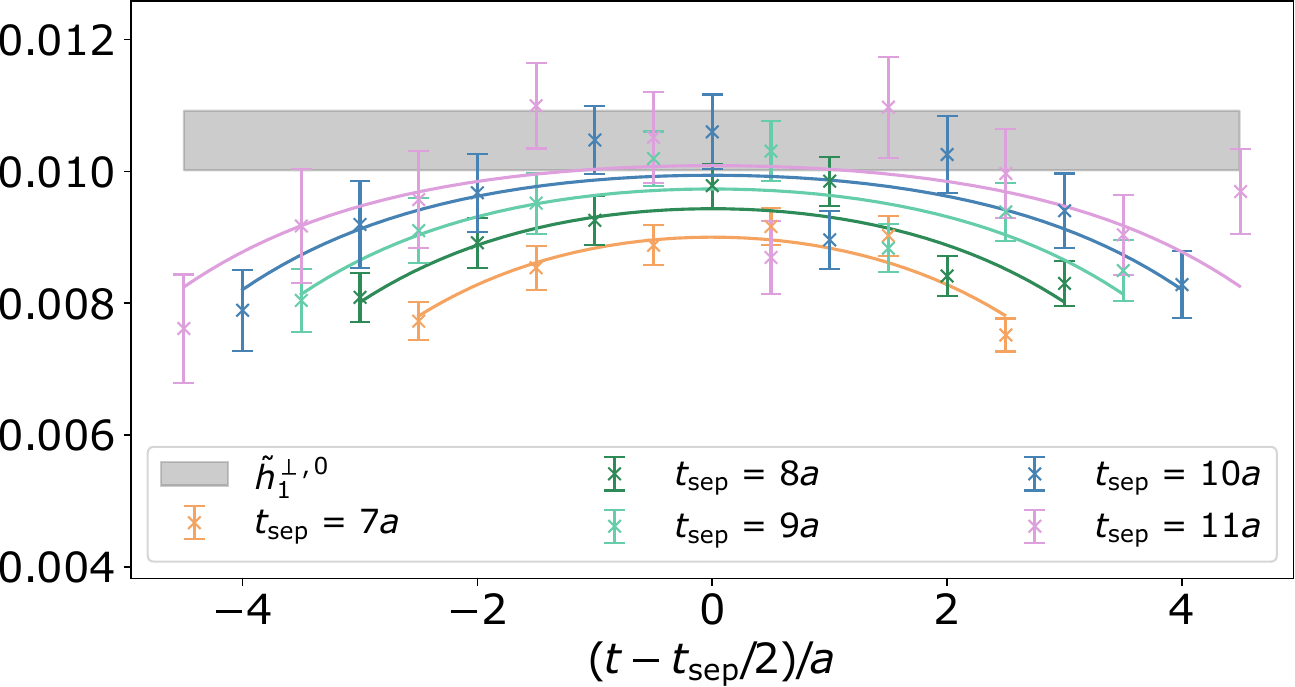}
\caption{H102, $P^z=0.91$~GeV, $z=a$, $b_{\perp}=3a$}
\label{ratio_fit_H102-1}
\end{subfigure}
\hfill
\begin{subfigure}{0.32\textwidth}
\includegraphics[width=\textwidth]{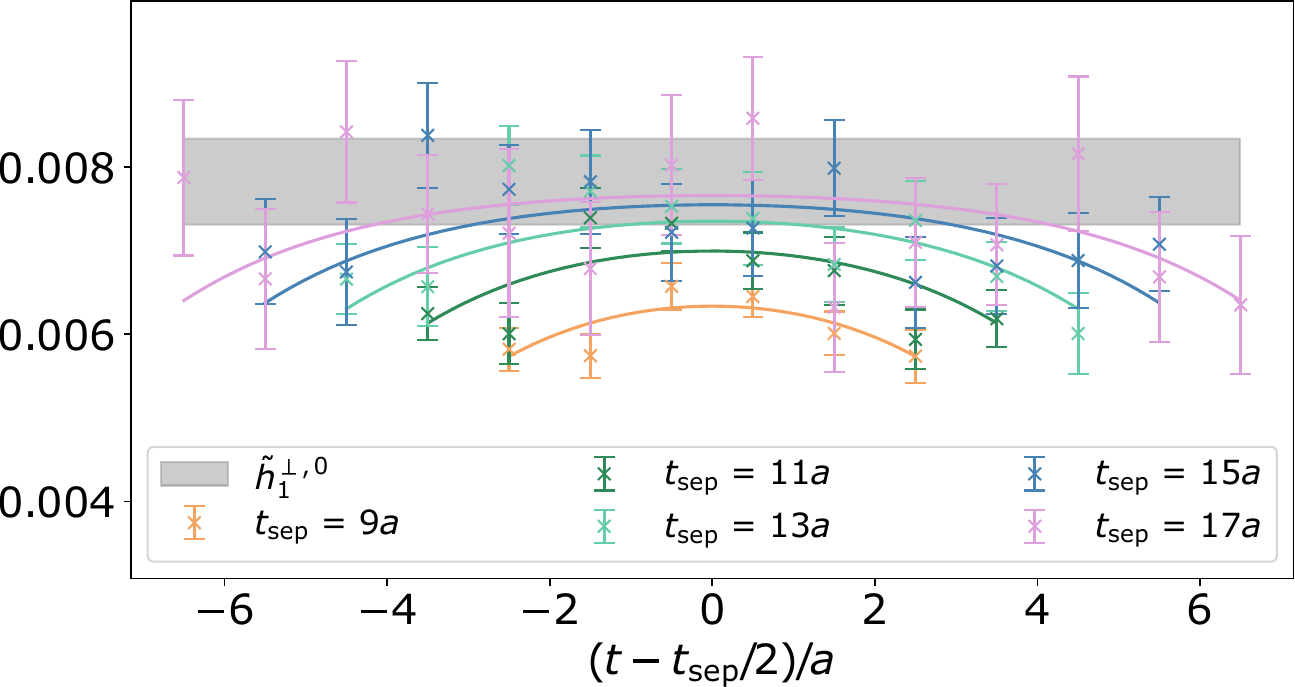}
\caption{N203, $P^z=0.81$~GeV, $z=a$, $b_{\perp}=3a$}
\label{fig:ratio_fit_N203-1}
\end{subfigure}\\
\hfill
\begin{subfigure}{0.32\textwidth}
\includegraphics[width=\textwidth]{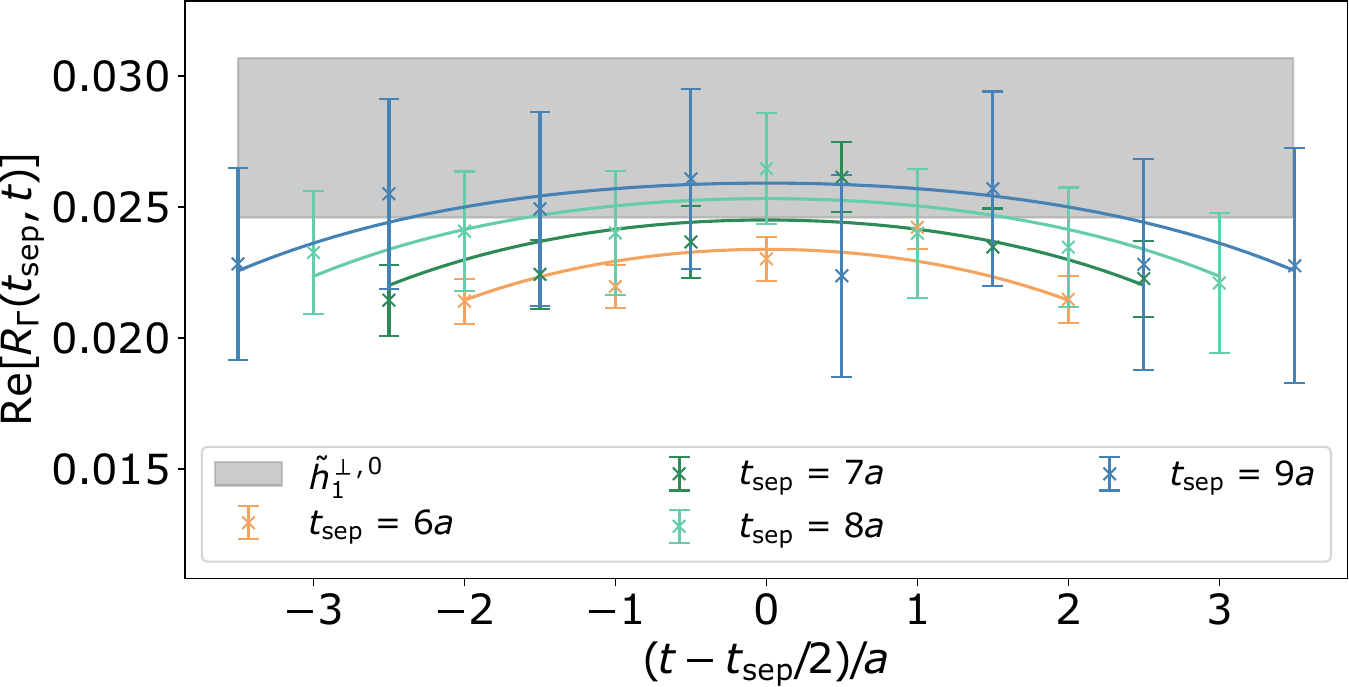}
\caption{X650, $P^z=1.32$~GeV, $z=a$, $b_{\perp}=3a$}
\label{ratio_fit_X650-2}
\end{subfigure}
\begin{subfigure}{0.32\textwidth}
\includegraphics[width=\textwidth]{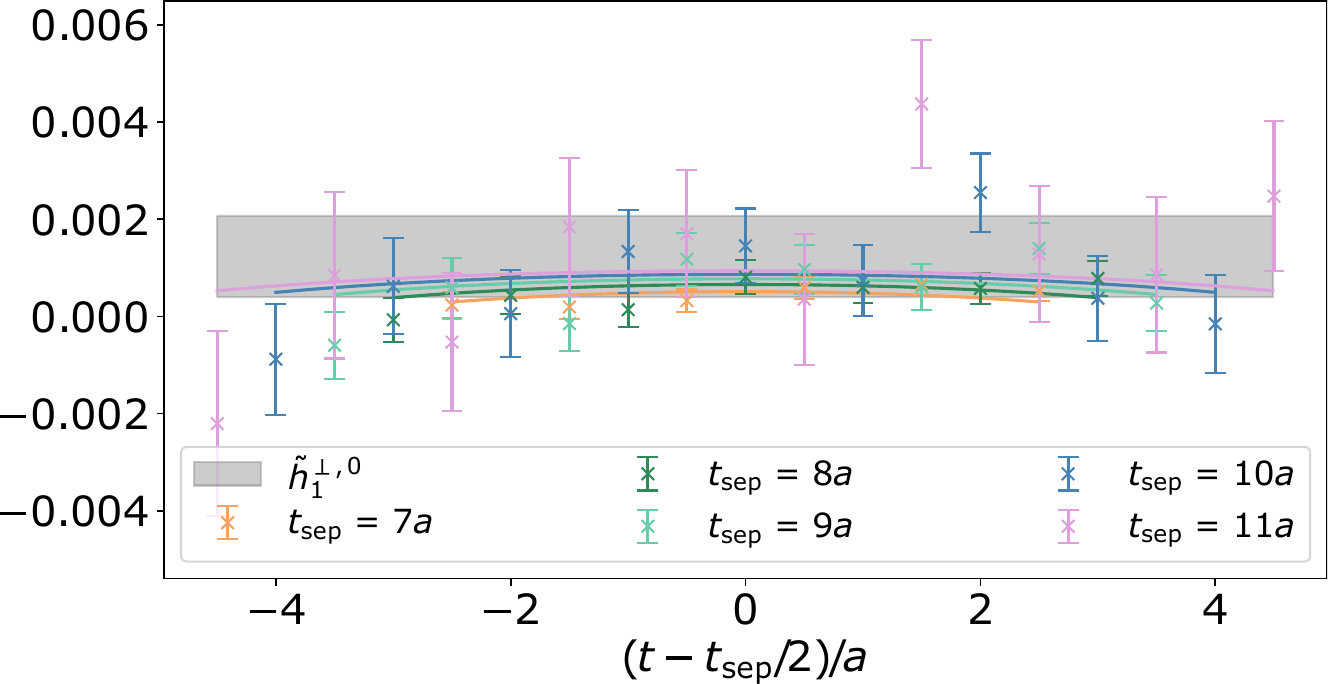}
\caption{H102, $P^z=1.37$~GeV, $z=7a$, $b_{\perp}=3a$}
\label{ratio_fit_H102-2}
\end{subfigure}
\hfill
\begin{subfigure}{0.32\textwidth}
\includegraphics[width=\textwidth]{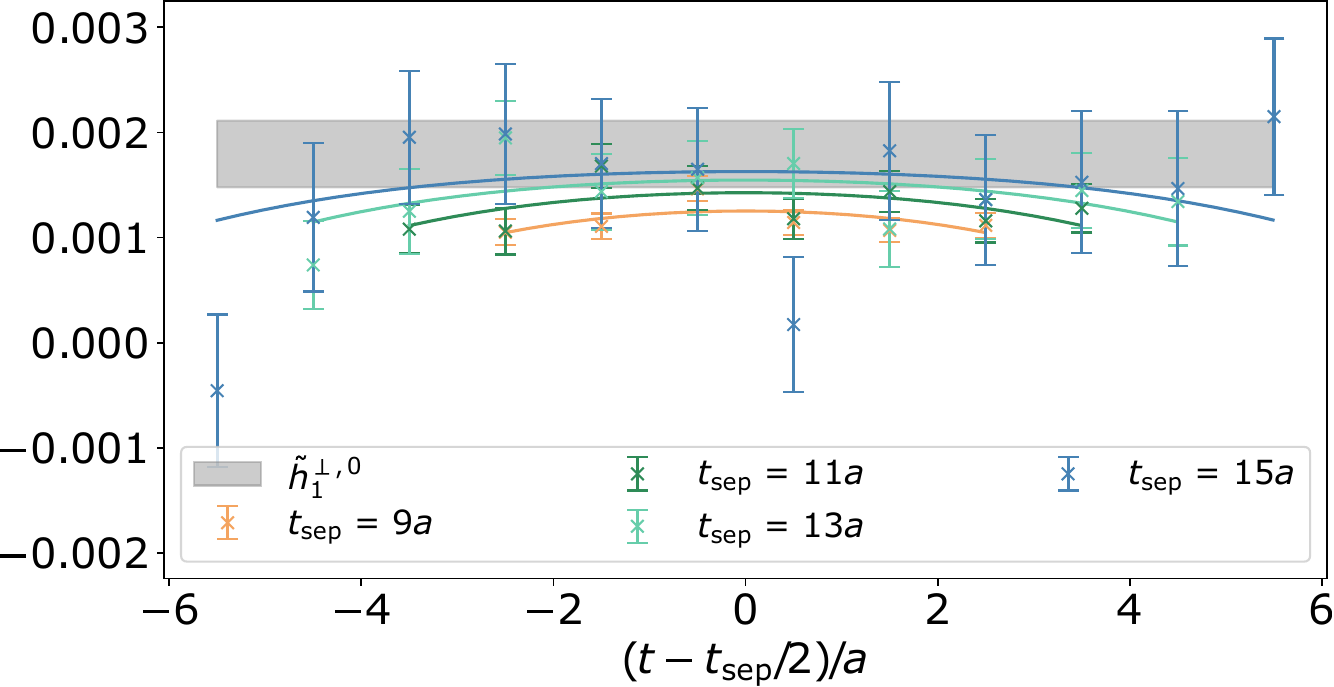}
\caption{N203, $P^z=1.21$~GeV, $z=5a$, $b_{\perp}=3a$}
\label{ratio_fit_N203-2}
\end{subfigure}\\
\hfill
\begin{subfigure}{0.32\textwidth}
\includegraphics[width=\textwidth]{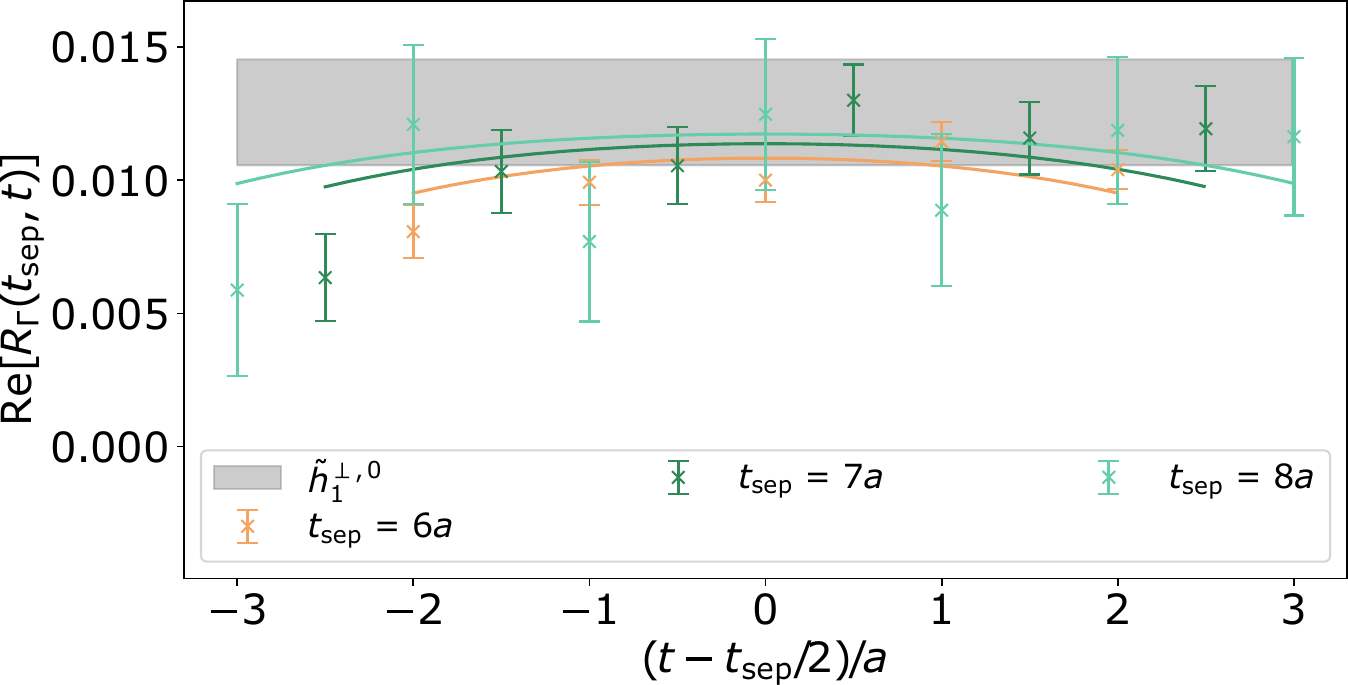}
\caption{X650, $P^z=1.58$~GeV, $z=3a$, $b_{\perp}=3a$}
\label{ratio_fit_X650-3}
\end{subfigure}
\hfill
\begin{subfigure}{0.32\textwidth}
\includegraphics[width=\textwidth]{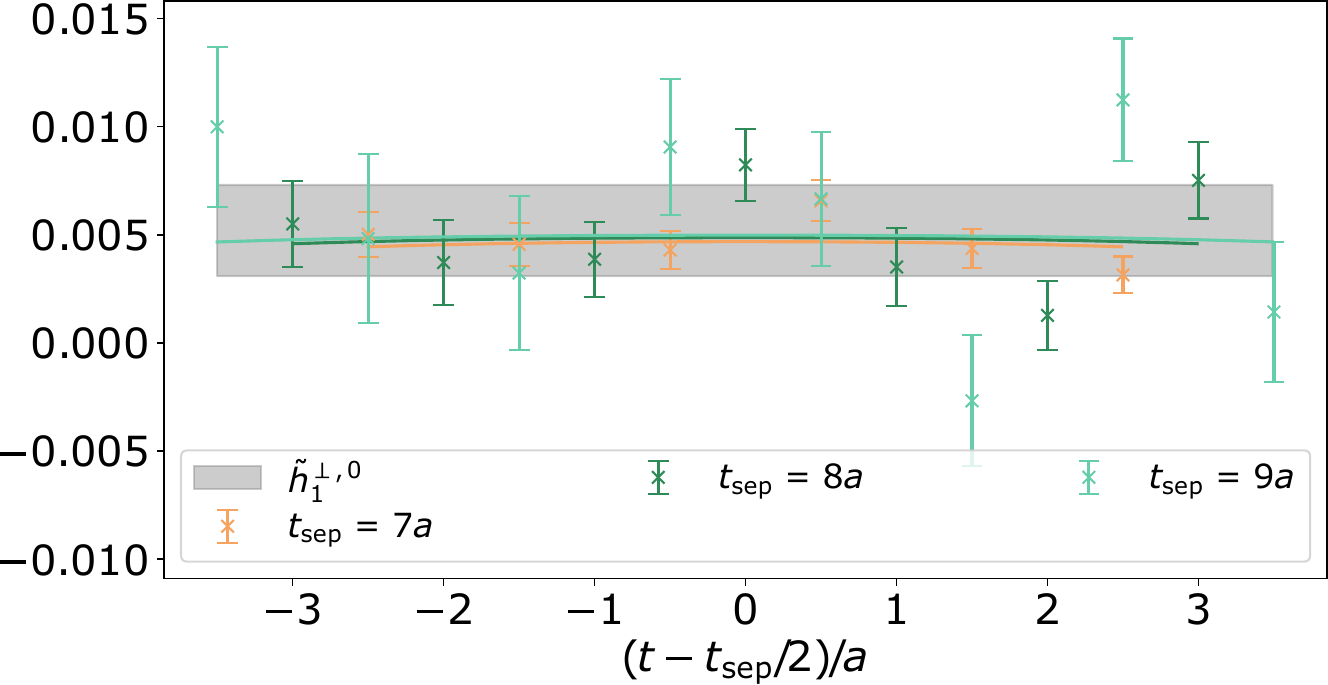}
\caption{H102, $P^z=1.82$~GeV, $z=a$, $b_{\perp}=3a$}
\label{ratio_fit_H102-3}
\end{subfigure}
\begin{subfigure}{0.32\textwidth}
\includegraphics[width=\textwidth]{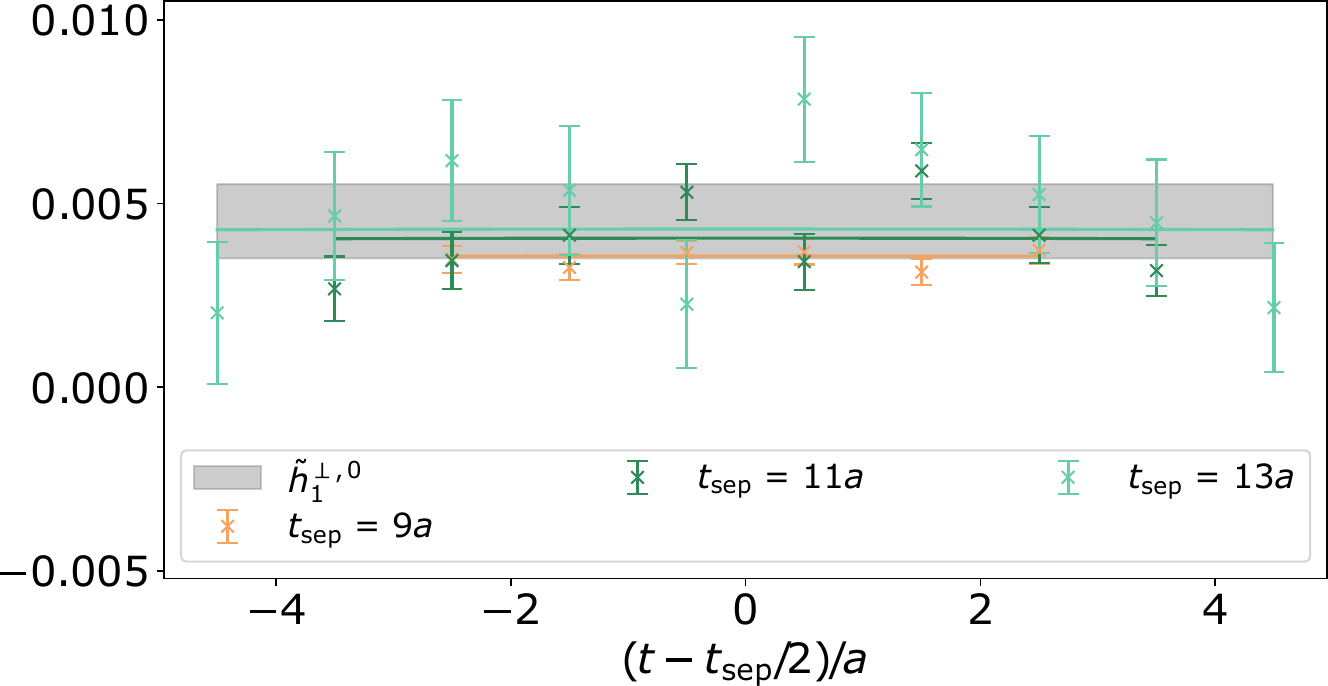}
\caption{N203, $P^z=1.61$~GeV, $z=a$, $b_{\perp}=3a$}
\label{ratio_fit_N203-3}
\end{subfigure}\\
\hfill
\begin{subfigure}{0.32\textwidth}
\includegraphics[width=\textwidth]{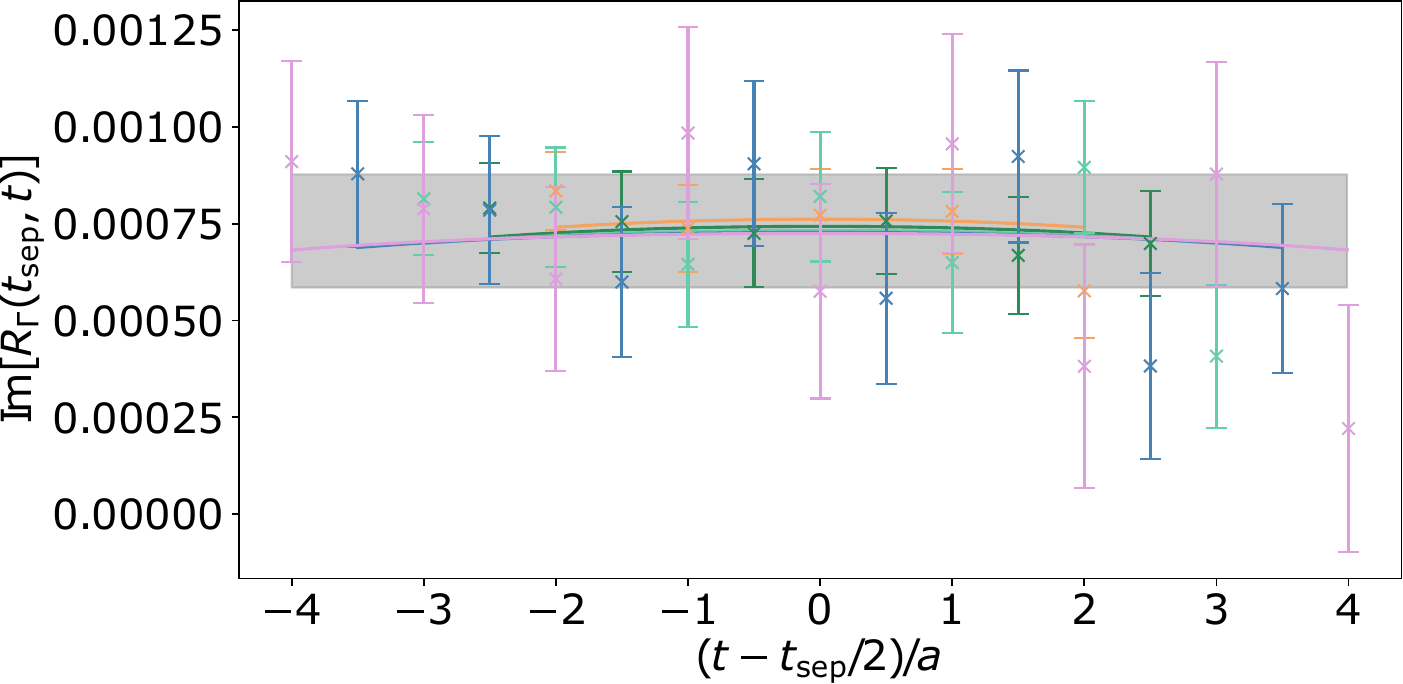}
\caption{X650, $P^z=0.79$~GeV, $z=a$, $b_{\perp}=3a$}
\label{ratio_fit_X650-4}
\end{subfigure}
\hfill
\begin{subfigure}{0.32\textwidth}
\includegraphics[width=\textwidth]{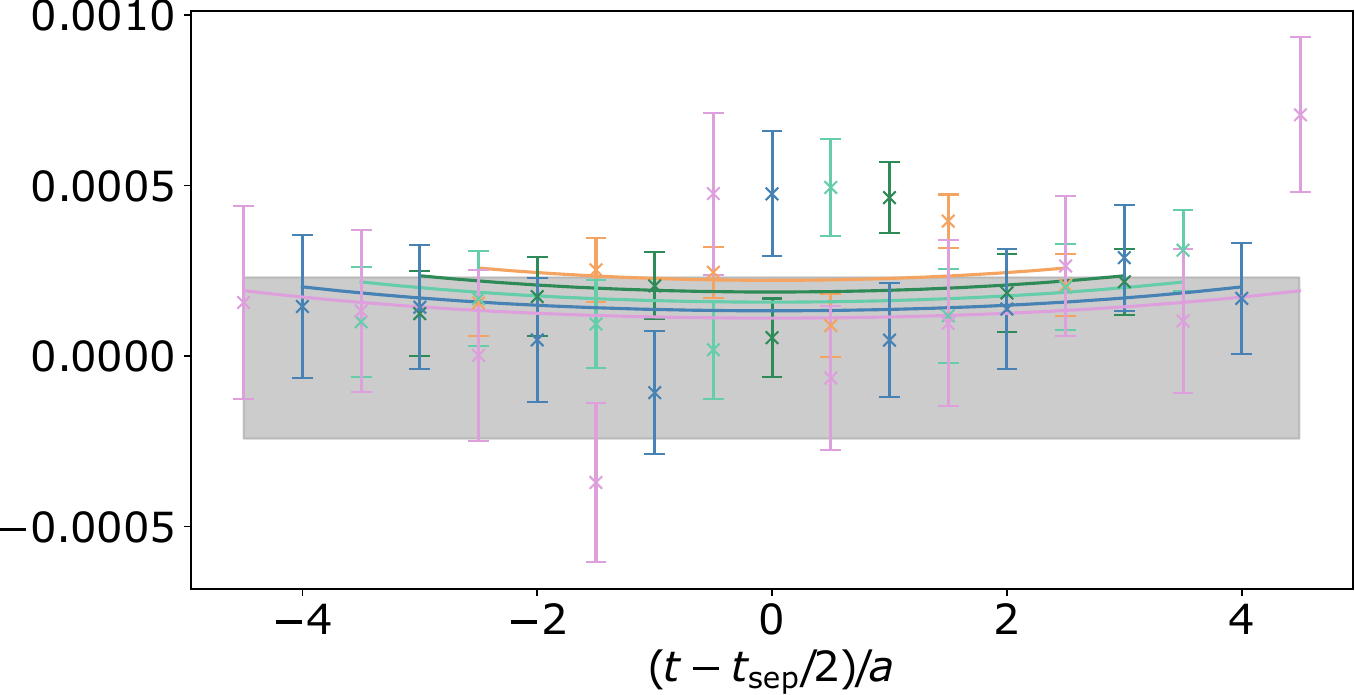}
\caption{H102, $P^z=0.91$~GeV, $z=a$, $b_{\perp}=3a$}
\label{ratio_fit_H102-4}
\end{subfigure}
\hfill
\begin{subfigure}{0.32\textwidth}
\includegraphics[width=\textwidth]{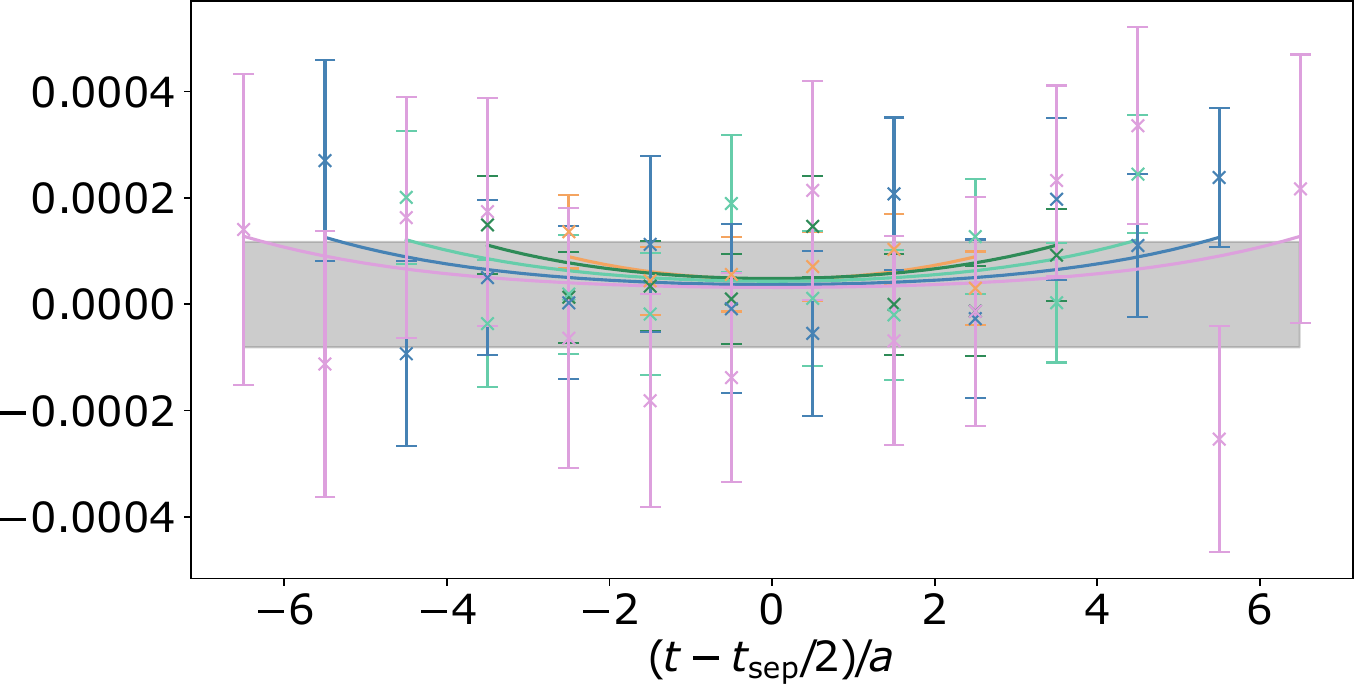}
\caption{N203, $P^z=0.81$~GeV, $z=a$, $b_{\perp}=3a$}
\label{ratio_fit_N203-4}
\end{subfigure}
\captionsetup{justification=raggedright,singlelinecheck=false}
\caption{Demonstration of fitting the correlation function for $b_{\perp}=3a$ for different ensembles, momenta and $z$. The ground state matrix element obtained by the fitting is shown as the gray band.} 
\label{fig:joint-fit}
\end{figure*}

In Fig.~\ref{fig:joint-fit}, we show, as a demonstrative example, the joint fit results for the ratio of three- and two-point functions for $b_{\perp}=3a$ for different ensembles, momenta, and $z$. As can be seen from the plot, the imaginary part of the matrix element is at least an order of magnitude smaller than the real part, and consistent with zero within errors in most cases.

\begin{figure}[thbp]
\includegraphics[width=.45\textwidth]{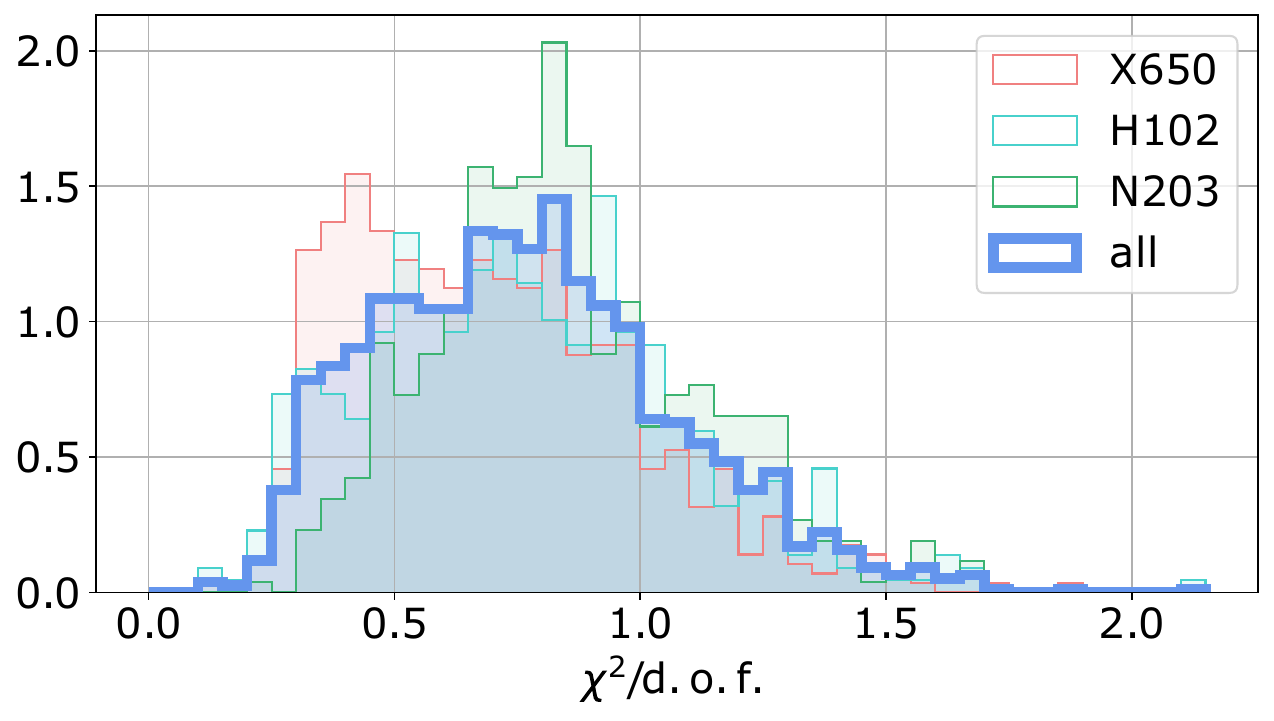}
\caption{Histograms of $\chi^2/{\text{d.o.f.}}$ of all fits to extract the bare ground state matrix element $\tilde{f}_\Gamma^0(z,b_{\perp},P^z,1/a)$. The distributions are also shown for the ensembles individually and are normalized to 1.} 
\label{fig:histogram_chi2_ovr_dof_ratio_fits}
\end{figure}

To illustrate the quality of the correlated fits, Fig. \ref{fig:histogram_chi2_ovr_dof_ratio_fits} shows the histograms of $\chi^2/\text{d.o.f.}$ from all correlated joint fits that are performed to determine the bare ground state matrix element $\tilde{h}^0(z,b_{\perp},P^z,1/a)$. The distributions are normalized to 1.

\subsection{$L$-dependence of subtracted quasi-TMDPDF matrix elements}

\begin{figure}[thbp]
\includegraphics[width=.45\textwidth]{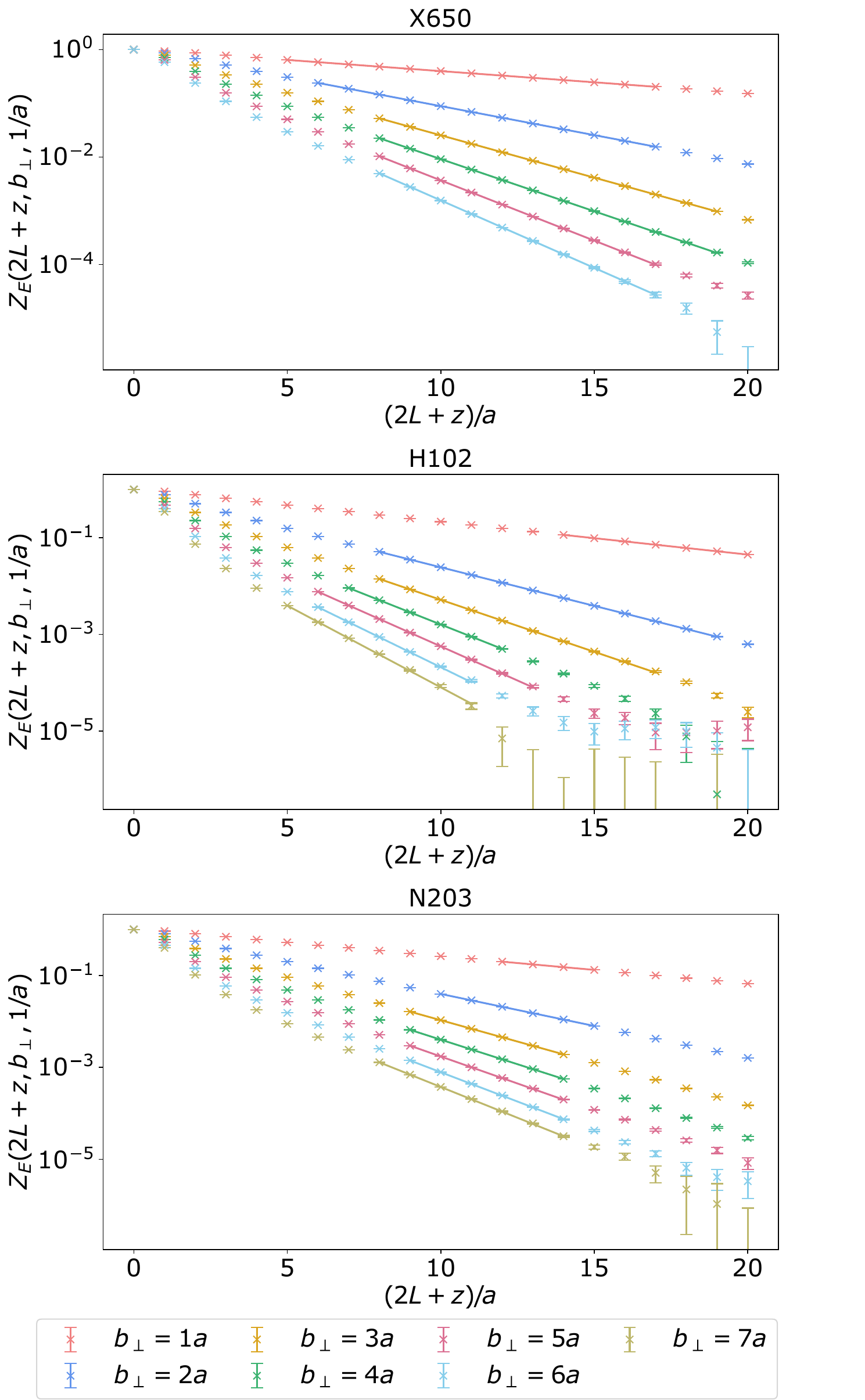}
\caption{The Wilson loop calculated and fitted on three different ensembles. The lines represent the fitted results.} 
\label{fig:WL}
\end{figure}

After dividing the bare matrix element by the square root of the Wilson loop, the singular dependence on $L$ is expected to cancel, see Eq. \eqref{eq:quasi_TMD}. To verify this, we calculate $Z_E$ for X650, H102 and N203 using all available gauge configurations. The signal-to-noise ratio of $Z_E(2L+z,b_{\perp},1/a)$ decays rapidly with increasing $2L+z$ and $b_{\perp}$, possibly even leading to negative central values. To decrease the uncertainty and avoid an ill-defined square root, we fit the Wilson loop with $Z_E(2L+z, b_{\perp},1/a)=c(b_{\perp,a}) e^{-V(b_{\perp},a)(2L+z)}$, with $V(b_{\perp},a)$ being the static QCD potential, and extrapolate to large values of $2L+z$, as was done in \cite{Zhang:2022xuw}. In Fig.~\ref{fig:WL}, we plot the Wilson loop calculated on three ensembles as well as the fitting result. The solid lines indicate the fitted ranges.

To investigate the $L$-dependence of the subtracted matrix elements, we calculate the three-point function for all gauge ensembles with various different values of $L$. To best utilize the available computing resources, smaller subsets of configurations and smaller momenta $P^z$ are used for this test. For H102, 100 configurations with eight measurements per configuration are analyzed, while for N203, 25 configurations with eight measurements per configuration are used. All available configurations are used for X650, with one measurement for each configuration.

\begin{figure}[thbp]
\includegraphics[width=.45\textwidth]{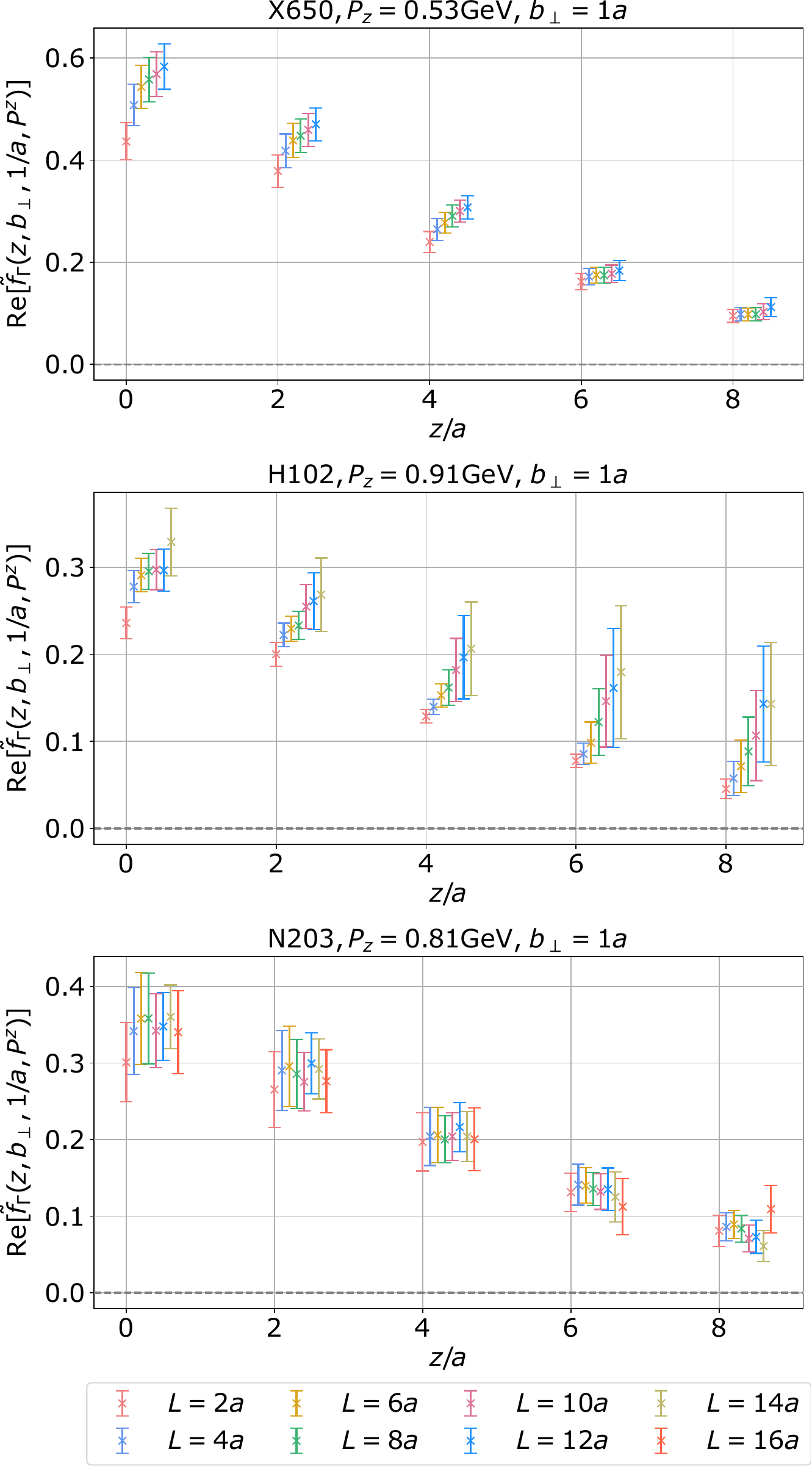}
\caption{The L dependence of the subtracted quasi-TMDPDF matrix elements on three ensembles.} 
\label{fig:L-indep}
\end{figure}

Fig.~\ref{fig:L-indep} shows the real parts of the subtracted quasi-TMDPDF matrix elements for X650 with $P^z=\SI{0.53}{\giga\electronvolt}$, for H102 with $P^z=\SI{0.91}{\giga\electronvolt}$ and for N203 with $P^z=\SI{0.81}{\giga\electronvolt}$, with $z=\{0,2,4,6,8\}a$. The minimum value $L_{\text{min}}=2a$ and step size $L_{\text{step}}=2a$ are the same for each ensemble, while the maximum value of $L$ is chosen as $L_{\text{max}}=12a$ for X650, $L_{\text{max}}=14a$ for H102 and $L_{\text{max}}=16a$ for N203. As is apparent from the figure, the $L$-dependence of the subtracted matrix elements is weak, a plateau is found for all ensembles and values of $z$. Since the signal becomes worse with increasing $L$, values of $L=8a$ for X650 and H102, as well as $L=10a$ for N203 are chosen for the calculation with full statistics.

\subsection{Renormalization}
The bare matrix elements are renormalized using the square root of the rectangular Euclidean Wilson loop $\sqrt{Z_E}$ and the logarithmic divergence factor $Z_O$. After constructing the subtracted quasi-TMDPDF as described above, the remaining logarithmic divergence can be removed by further dividing by a short distance matrix element at zero momentum, as discussed in Ref.~\cite{Zhang:2022xuw}.

\begin{figure*}[thbp]
\includegraphics[width=\textwidth]{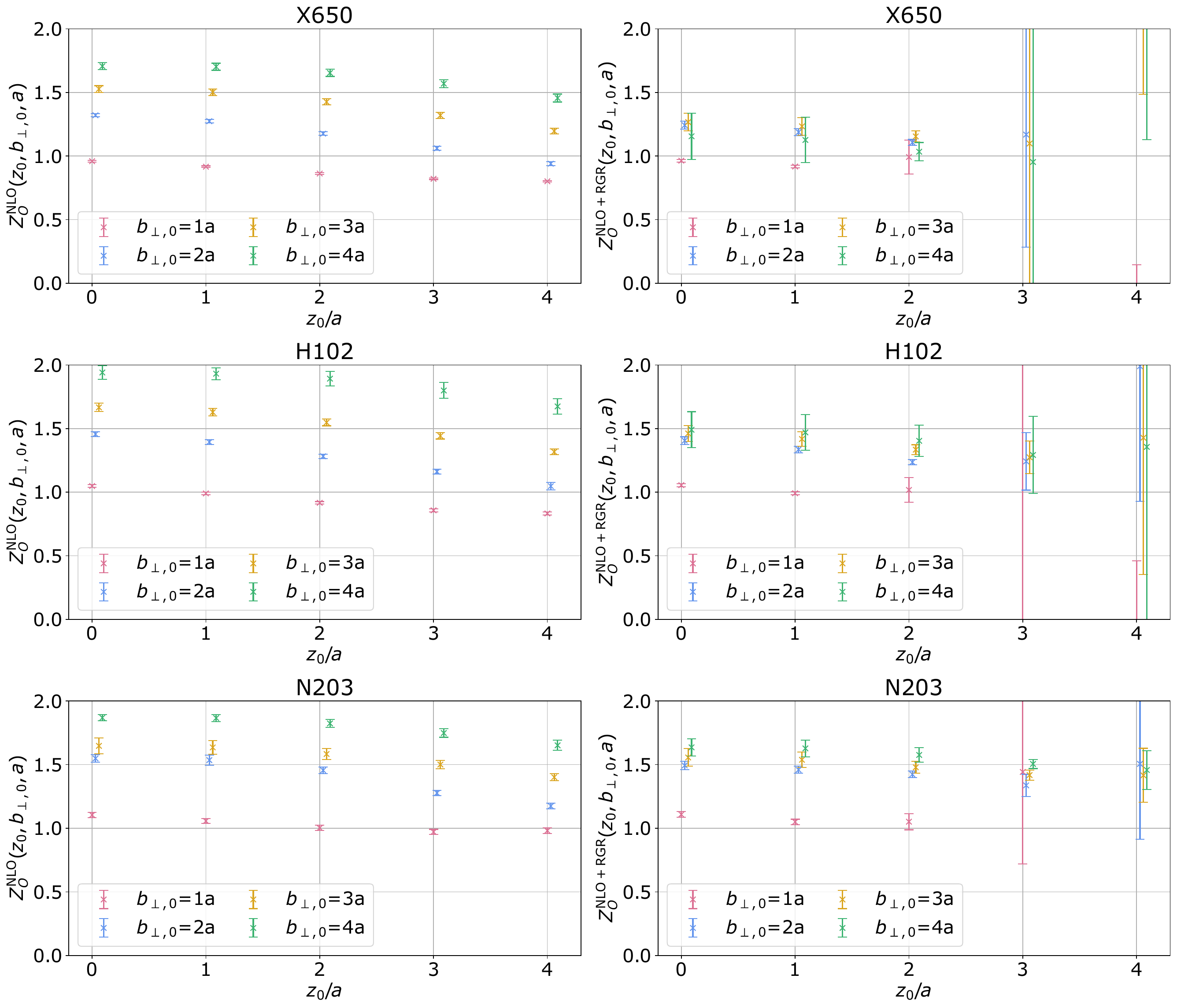}
\vspace{-1em}
\caption{The determination of the renormalization factor $Z_O$ without (labeled as NLO) and with including the RGR effect (labeled as NLO+RGR) on three ensembles. Windows of constant $Z_O$ can be found in the right panel.} 
\label{fig:ZO}
\end{figure*}

The renormalization factor $Z_O$ will introduce some residual dependence on $z_0, b_{\perp, 0}$ due to missing higher-order perturbative contributions. To reduce such dependence, we have carried out an RG resummation (RGR) for the perturbative $\overline{\rm MS}$ result with the physical scale being set to $\mu_0 = r \cdot 2 e^{-\gamma_E} / \sqrt{b_{\perp,0}^2 + z_0^2}$ with prefactor $r$ to be $\mathcal{O}(1)$. 
In Fig.~\ref{fig:ZO}, we show the comparison of the logarithmic divergence factors $Z_O$ at various distances $z_0, b_{\perp,0}$ calculated at NLO without (left panel) and with (right panel) RGR. The renormalization scale is $\mu=\SI{2}{\giga\electronvolt}$ and the physical scale in the resummation is varied from $r=0.8$ to $r=1.2$, leading to a systematic uncertainty which is included in the error bars of the right panel in the figure.

Windows of constant $Z_O$ are found at $b_{\perp,0}=\{2,3\}a$, $z_0=\{0,1,2\}a$ for X650, at $b_{\perp,0}=\{2,3\}a$, $z_0=\{0,1\}a$ for H102, and at $b_{\perp,0}=\{2,3\}a$, $z_0=\{0,1,2\}a$ for N203. Our final renormalization factors are obtained by averaging $Z_O$ over those regions. The resulting values for each ensemble are $Z_O^{\text{X650}}=1.199(16)(43)$, $Z_O^{\text{H102}}=1.405(20)(38)$ and $Z_O^{\text{N203}}=1.469(32)(17)$, with the first error being the statistical uncertainty estimated with bootstrap resampling and the second error arising from the scale variation in RGR.

\subsection{Large $\lambda$ extrapolation}

\begin{figure*}[thbp]
\includegraphics[width=\textwidth]{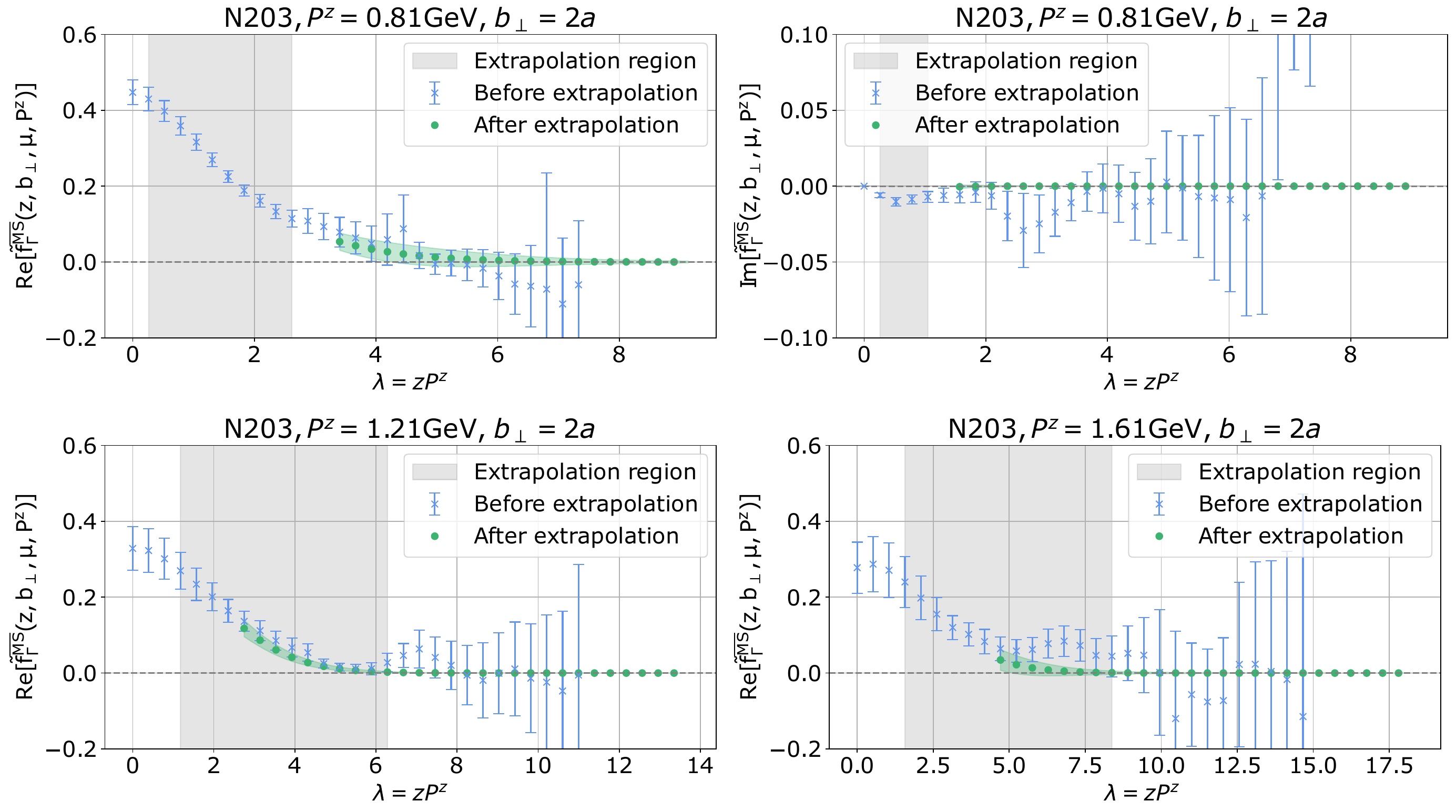}
\vspace{-1em}
\caption{The extrapolation of the renormalized quasi-TMDPDF matrix element in the large-$\lambda$ region. Here we take the data at $b_\perp=2a$ and $P^z=0.81$ GeV, $P^z=1.21$ GeV and $P^z=1.61$ GeV on N203 as an example.} 
\label{fig:extrap}
\end{figure*}

In order to Fourier transform the renormalized matrix elements in coordinate space to momentum space, we need the quasi-TMDPDF matrix elements at all quasi-LF distances $\lambda=z P^z$. However, the uncertainty grows quickly at large distances. Here we adopt the following extrapolation form~\cite{Ji:2020brr} at large $\lambda$ inspired by the analysis of collinear PDFs~\cite{Ji:2020brr}
\begin{align}
	  H^{\rm R}_{m}(\lambda) &= \Big[\frac{c_1}{(i\lambda)^a} + e^{-i\lambda}\frac{c_2}{(-i \lambda)^b}\Big]e^{-\lambda/\lambda_0},
	  \label{eq:extrap}
\end{align} 
where $c_1, c_2$ in the square bracket can depend on $b_\perp$, and the exponential term comes from the expectation that at finite momentum the correlation function has a finite correlation length (denoted as $\lambda_0$)~\cite{Ji:2020brr}, which becomes infinite when the momentum goes to infinity. The details of the extrapolation are expected to affect the final results in the endpoint $x$ regions where LaMET expansion breaks down~\cite{Ji:2020ect}. 

The $\lambda$ extrapolation is performed independently for each value of $b_{\perp}$ and $P^z$. Examples for extrapolated results are shown in Fig.~\ref{fig:extrap} for the real part of the renormalized matrix elements for N203, $b_{\perp}=2a$, $P^z=\{0.81,1.21,1.61\}\si{\giga\electronvolt}$. For $P^z=\SI{0.81}{\giga\electronvolt}$, also the corresponding imaginary part is depicted. The data in the grey extrapolation region is used during the fitting process, and the grey region is determined by minimizing the $\chi^2/\text{d.o.f.}$ of the correlated fit. The result after extrapolation is shown in green. The region where the extrapolated result is used instead of the original data is chosen as small as possible, but large enough to ensure valid results after the Fourier transformation. The results of the extrapolation agree with the lattice data in the region of moderate $\lambda$ and give smooth curves with much reduced errors for large $\lambda$.

After renormalization and extrapolation, we can do a Fourier transform to $(x, b_\perp)$ space. The imaginary part of the coordinate space distributions, which is zero within errors in most cases, is treated as part of the systematic uncertainties.

\subsection{Linear interpolation in $b_{\perp}$ of the Collins-Soper kernel and intrinsic soft function}

\begin{figure*}[thbp]
\includegraphics[width=\textwidth]{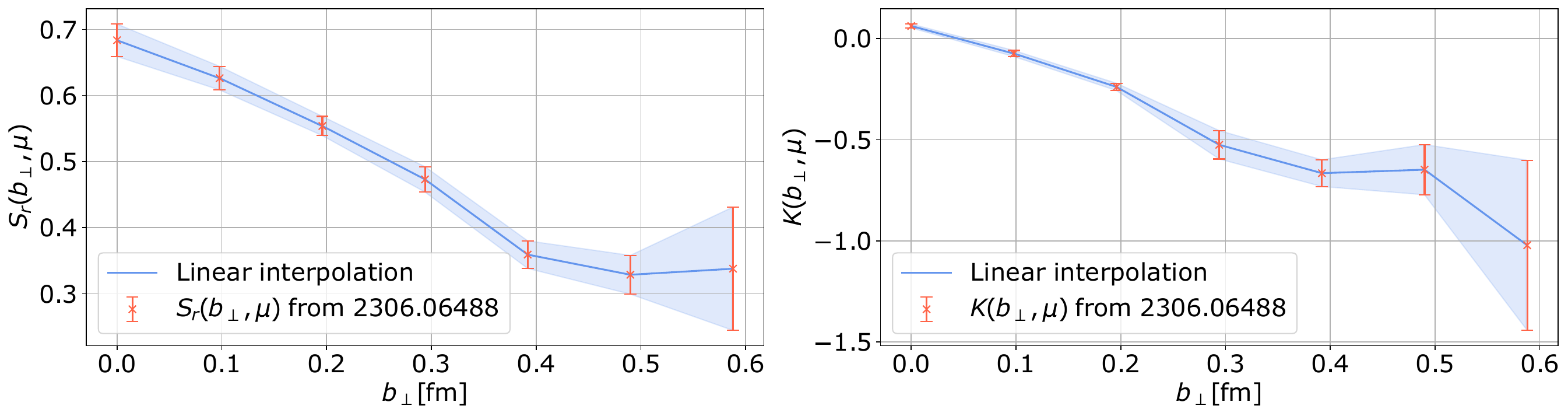}
\vspace{-1em}
\caption{The $b_\perp$ interpolation of the intrinsic soft function and the Collins-Soper kernel calculated on X650 so that they can be used for H102 and N203.} 
\label{fig:interp}
\end{figure*}

To apply the perturbative matching for different lattice spacings, the Collins-Soper kernel $K(b_{\perp},\mu)$ and intrinsic soft function $S_I(b_{\perp},\mu)$ at different values of $b_{\perp}$ are necessary. Both quantities have been calculated on the ensemble X650 at $\mu=\SI{2}{\giga\electronvolt}$ in \cite{LatticePartonLPC:2023pdv}, but are not yet available on H102 and N203. As a makeshift we perform a linear interpolation so that they can be used for the other two ensembles, with the results shown in Fig.~\ref{fig:interp}. 

Since the results for $K(b_{\perp},\mu)$ and $S_I(b_{\perp},\mu)$ from \cite{LatticePartonLPC:2023pdv} are not available for the exact same bootstrap samples that are used in our analysis, we use error propagation to incorporate the uncertainties of the Collins-Soper kernel and intrinsic soft function in our results after matching. The uncertainties that we use in the error propagation are obtained by linearly interpolating the statistical errors from \cite{LatticePartonLPC:2023pdv} to the desired values of $b_{\perp}$ needed for our analysis.

\subsection{Impact of perturbative matching and resummation effect}

\begin{figure*}[thbp]
\includegraphics[width=\textwidth]{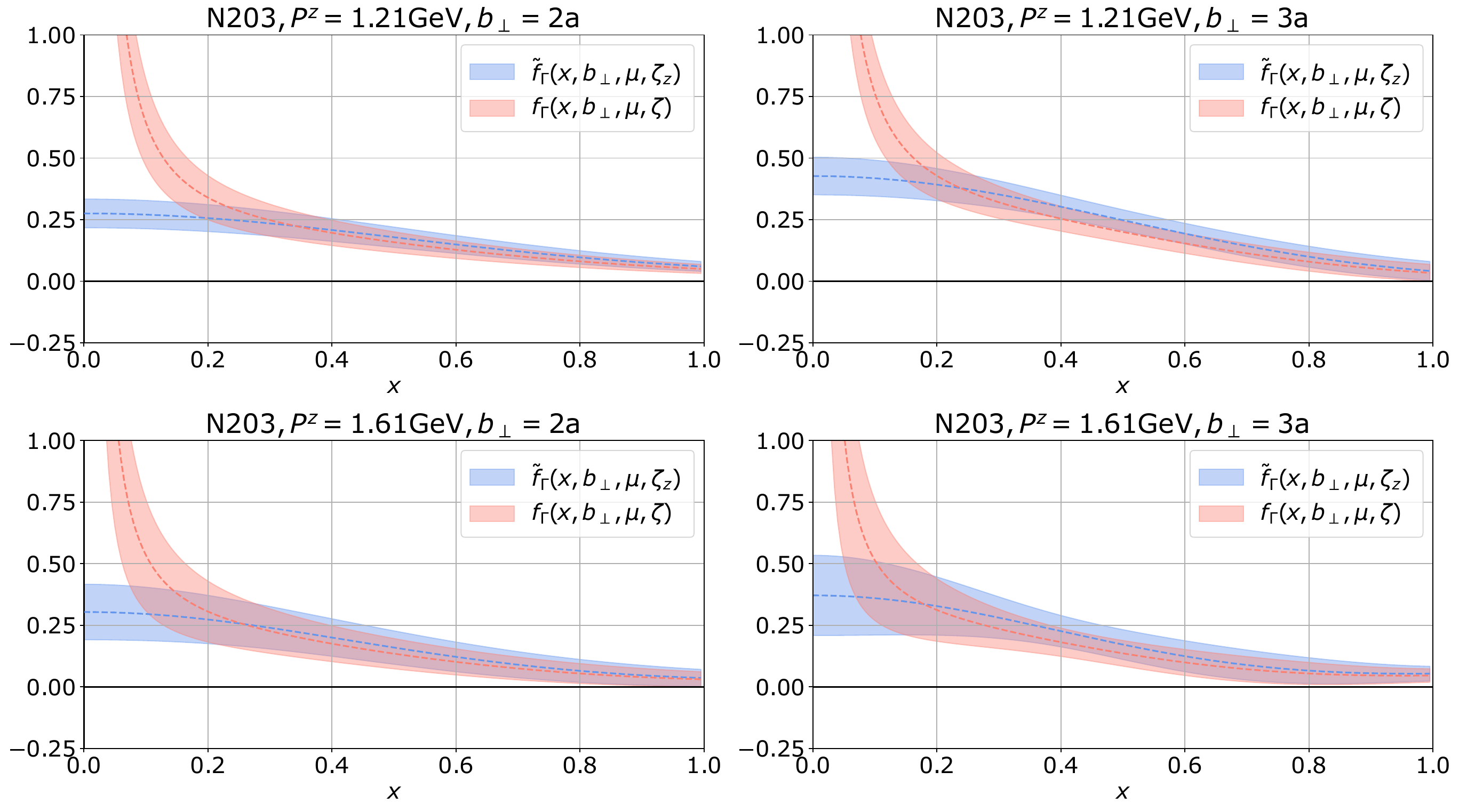}
\vspace{-1em}
\caption{Impact of the perturbative matching at NNLO. We take the data of the renormalized quasi-TMDPDF $\tilde{f}_{\Gamma}(x,b_{\perp},\mu,\zeta_z)$ and the TMDPDF $f_{\Gamma}(x,b_{\perp},\mu,\zeta)$ at $b_\perp=2a, 3a$ and $P^z=\{1.21, 1.61\}$ GeV on N203 as an example.} 
\label{fig:pertmat}
\end{figure*}

In Fig.~\ref{fig:pertmat}, we plot a comparison between the results before and after applying the perturbative matching at NNLO with renormalization scale $\mu=\SI{2}{\giga\electronvolt}$ and rapidity scale $\zeta=\SI{4}{\giga\electronvolt\squared}$. The data of N203, $b_{\perp}=2a,3a$ and $P^z=\{1.21,1.61\}\si{\giga\electronvolt}$ are shown as an example. The errors of the quasi-TMDPDF in the figure contain statistical errors as well as the error from the scale variation in RGR when determining $Z_O$. The errors of the TMDPDF additionally contain statistical uncertainties from the intrinsic soft function and Collins-Soper kernel. In the case of $P^z=\SI{1.61}{\giga\electronvolt}$, $b_{\perp}=3a$, the systematic uncertainty from the imaginary part is taken into account as described above. As can be seen from the figure, the matching mainly affects the small $x$ region.

\begin{figure}[thbp]
\includegraphics[width=.45\textwidth]{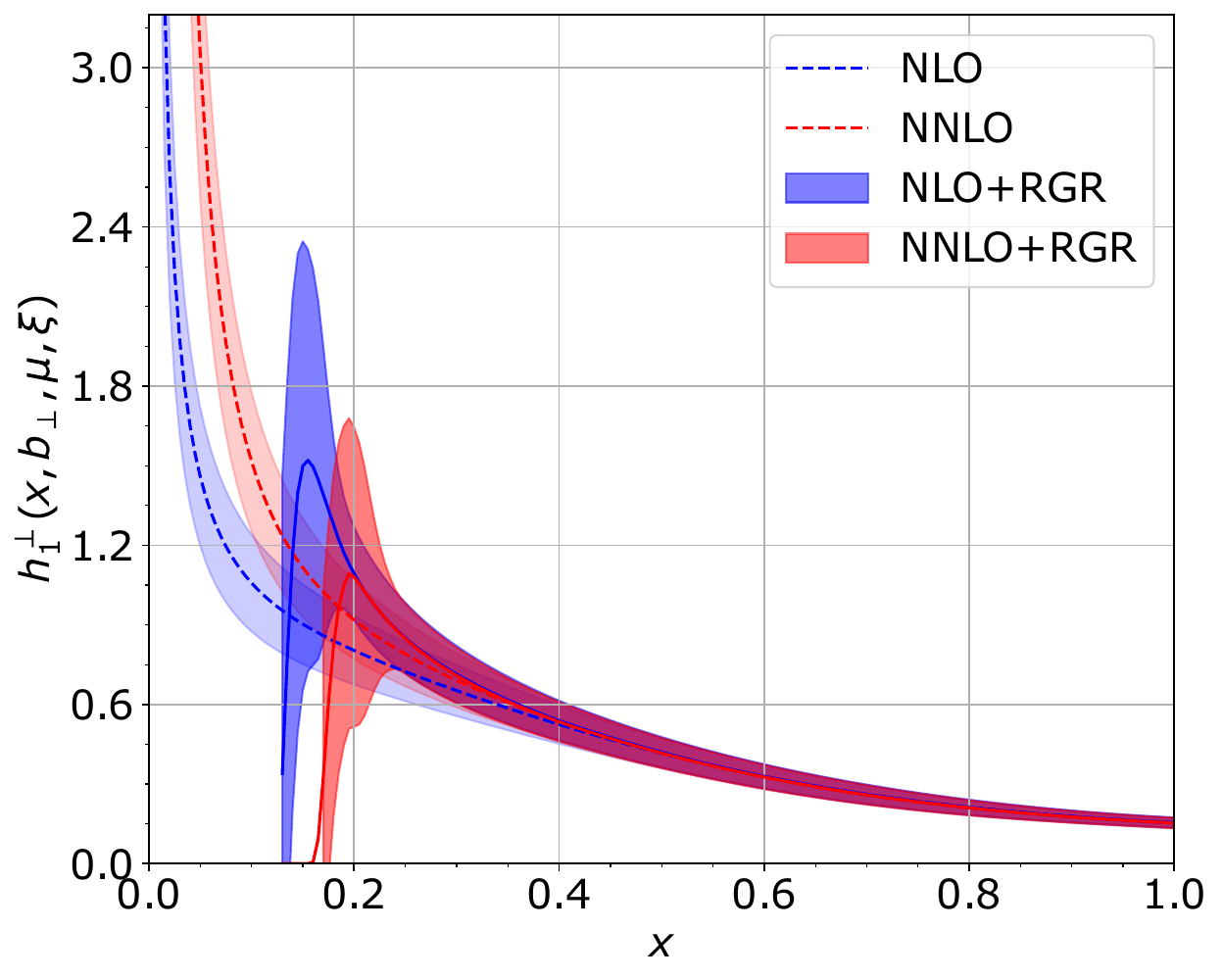}
\caption{Boer-Mulders function $h_1^{\perp}(x,b_{\perp},\mu,\zeta)$ for X650 with $P^z=\SI{1.58}{\giga\electronvolt}$ and $b_{\perp}=2a$, obtained with NLO (blue) and NNLO (red) matching kernel with and without RGR at scales $\mu=\SI{2}{\giga\electronvolt}$ and $\zeta=\SI{4}{\giga\electronvolt\squared}$. The error band includes statistical errors as well as the systematic error from scale variation in the resummation.} 
\label{fig:RGR_matching_NLO_NNLO}
\end{figure}

Fig. \ref{fig:RGR_matching_NLO_NNLO} shows the effects of using RGR of the perturbative matching kernel at NLO and NNLO for the example of X650, $P^z=\SI{1.58}{\giga\electronvolt}$ and $b_{\perp}=2a$. Statistical errors as well as the systematic error from varying the scale during the resummation are included in the error bands. The significant difference between the fixed order result and the RGR result in the small $x$ region clearly indicates that LaMET prediction is unreliable in this region where power corrections also become important and should be taken into account. 

\section{Numerical results and discussions}
\label{SEC:numres}

\subsection{Dependence of momentum space distributions on $P^z$}
\label{subsec:p_dependence}

\begin{figure}[thbp]
\includegraphics[width=.45\textwidth]{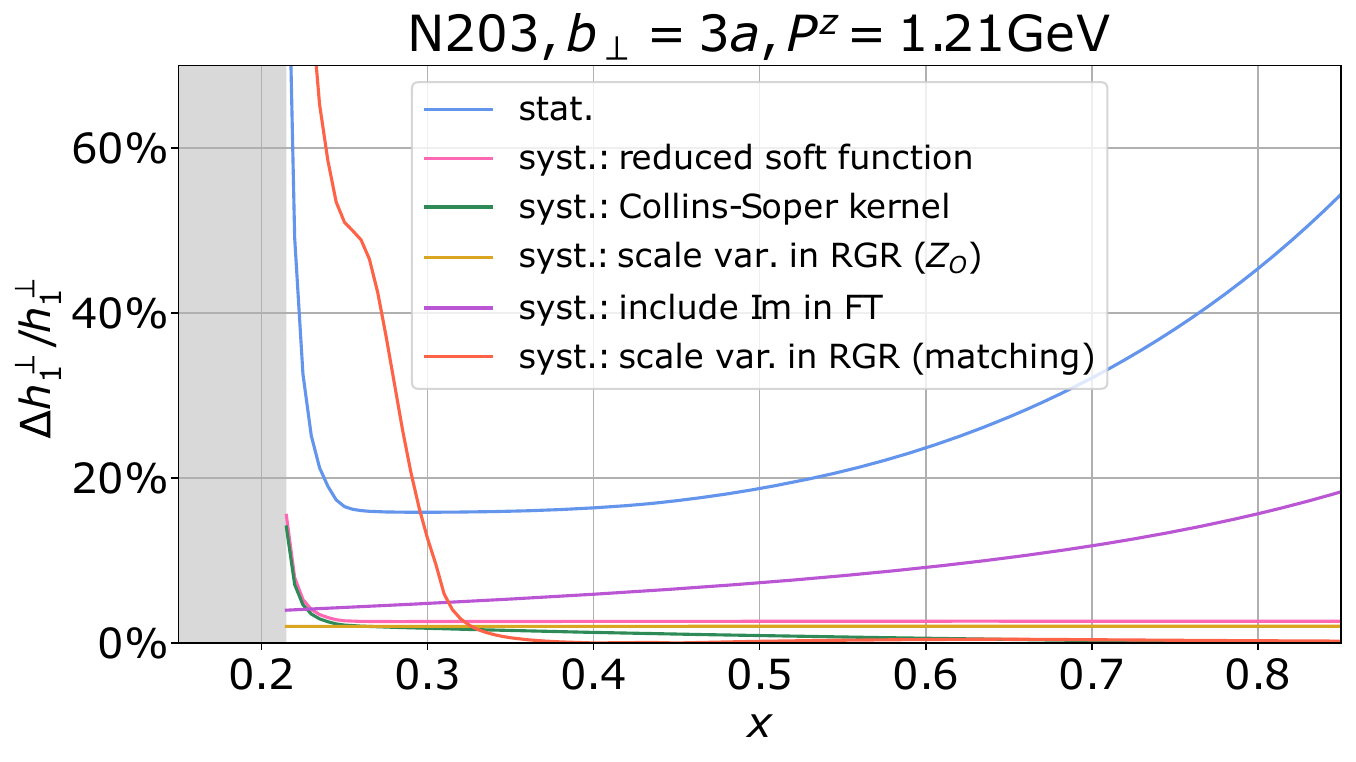}
\caption{Sources of systematic errors. We have taken the N203 result with $P^z=\SI{1.21}{\giga\electronvolt}$ and $b_\perp=3a$ after matching at NNLO with RGR as an example.} 
\label{fig:error}
\end{figure}

\begin{figure*}[thbp]
\includegraphics[width=\textwidth]{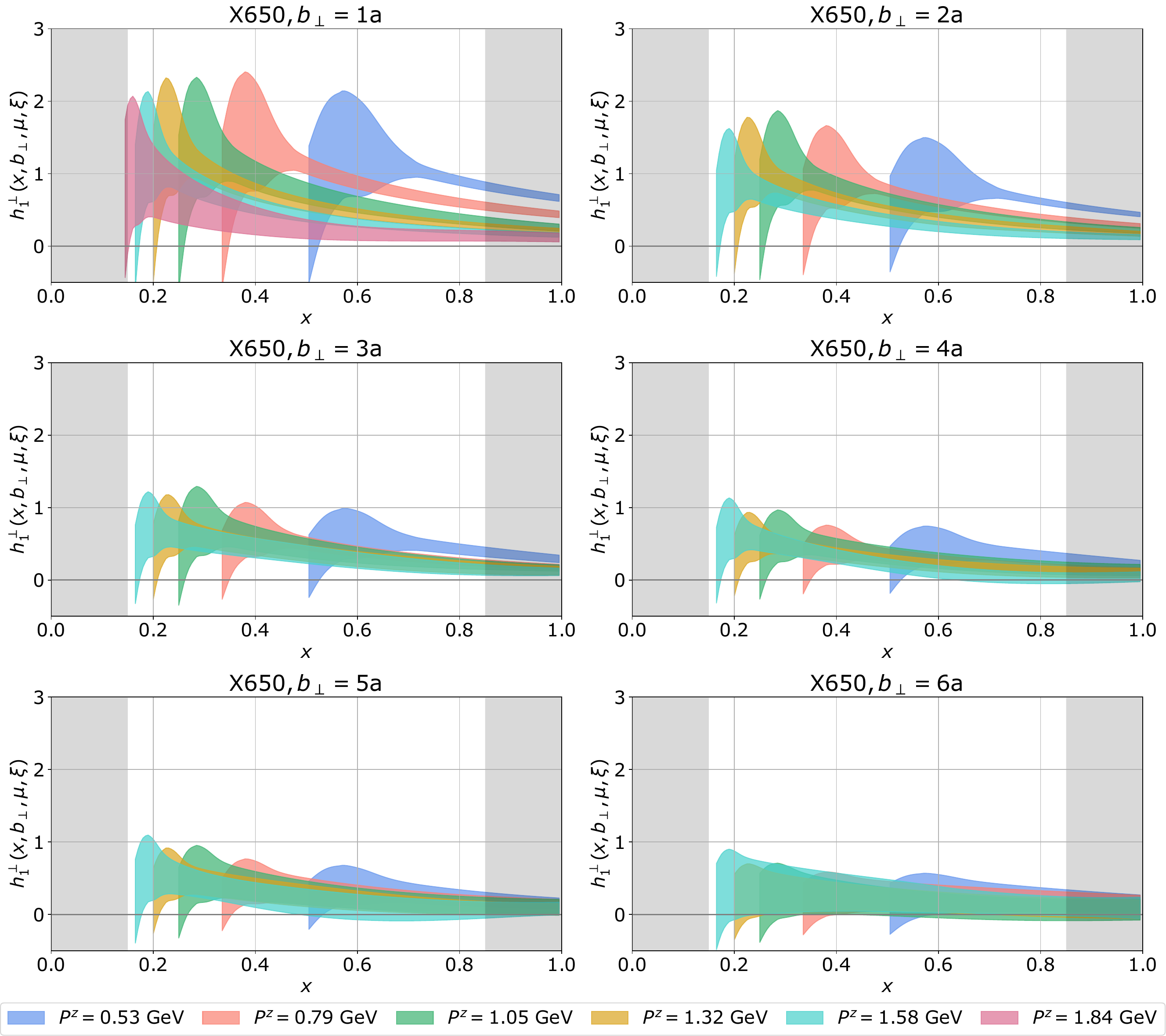}
\vspace{-1em}
\caption{Results on the Pion Boer-Mulders function of X650, showing the dependence on $P^z$ after matching at NNLO including RGR.} 
\label{fig:p_dependence_light_cone_result_X650}
\end{figure*}

Fig.~\ref{fig:p_dependence_light_cone_result_X650} shows, as an example, the pion Boer-Mulders function obtained on X650 for different pion momenta $P^z$ after matching at NNLO with RGR. The largest momentum $P^z=\SI{1.84}{\giga\electronvolt}$ for X650 is only included in the figure for $b_{\perp}=1a$ due to large error bars hindering visibility. Overall, the results exhibit good convergence with increasing momenta. 
The results for H102 and N203 exhibit a similar behavior as in Fig.~\ref{fig:p_dependence_light_cone_result_X650}.

The results after RGR can not be determined for small $x$, since the resummation breaks down due to a Landau pole. As can be seen from  Fig.~\ref{fig:p_dependence_light_cone_result_X650}, the $x$ range for which the corrections are under control gets larger for increasing $P^z$ and the resummation breaks down at $x \approx 0.15$ for the largest momentum of $P^z \approx \SI{1.84}{\giga\electronvolt}$ used for X650. Hence, we estimate the unreliable region to be $x \in [0,0.15]$ and $x \in [0.85,1]$ and shade this area in grey in Fig.~\ref{fig:p_dependence_light_cone_result_X650}. 

\subsection{Estimation of systematic uncertainties}
\label{subsec:systematic_uncertainties}

The error bands in Fig.~\ref{fig:p_dependence_light_cone_result_X650} include both statistical and systematic uncertainties, where the latter have five different sources, as illustrated in Fig.~\ref{fig:error}. 
The first is from the intrinsic soft function. 
The second comes from the Collins-Soper kernel. The third is from scale variation in the RGR for the renormalization factor $Z_O$. The fourth source of systematic uncertainty is the error due to neglecting the imaginary part of the matrix element. Finally, we also include the uncertainty from scale variation during RGR of the matching kernel.

\subsection{Fit of $b_{\perp}$-dependence}
\label{subsec:b_dependence}

In order to compare the results from ensembles with different lattice spacings, the dependence of the Boer-Mulders function on $b_{\perp}$ is fitted with
\begin{equation}
h_1^{\perp}(x,b_{\perp},\mu,\zeta) = c_1 (x,P^z,a) e^{-c_2 (x,P^z,a) \cdot b_{\perp}^2},
\label{eq:Boer-Mulders_fit_form_b_dep_gauss}
\end{equation}
which is guided by the global fits of the Pavia group, e.g., in \cite{Delcarro:2018lbr}. The fit form in Eq. \eqref{eq:Boer-Mulders_fit_form_b_dep_gauss} is simplified compared to the one used in \cite{Delcarro:2018lbr}, since the quality of the data and the number of different $b_{\perp}$ in this analysis do not allow for a fit with more parameters. The fit is performed for each ensemble and momentum individually, as well as for various values of $x$ with a step size of 0.005.

\begin{figure}[thbp]
\includegraphics[width=.45\textwidth]{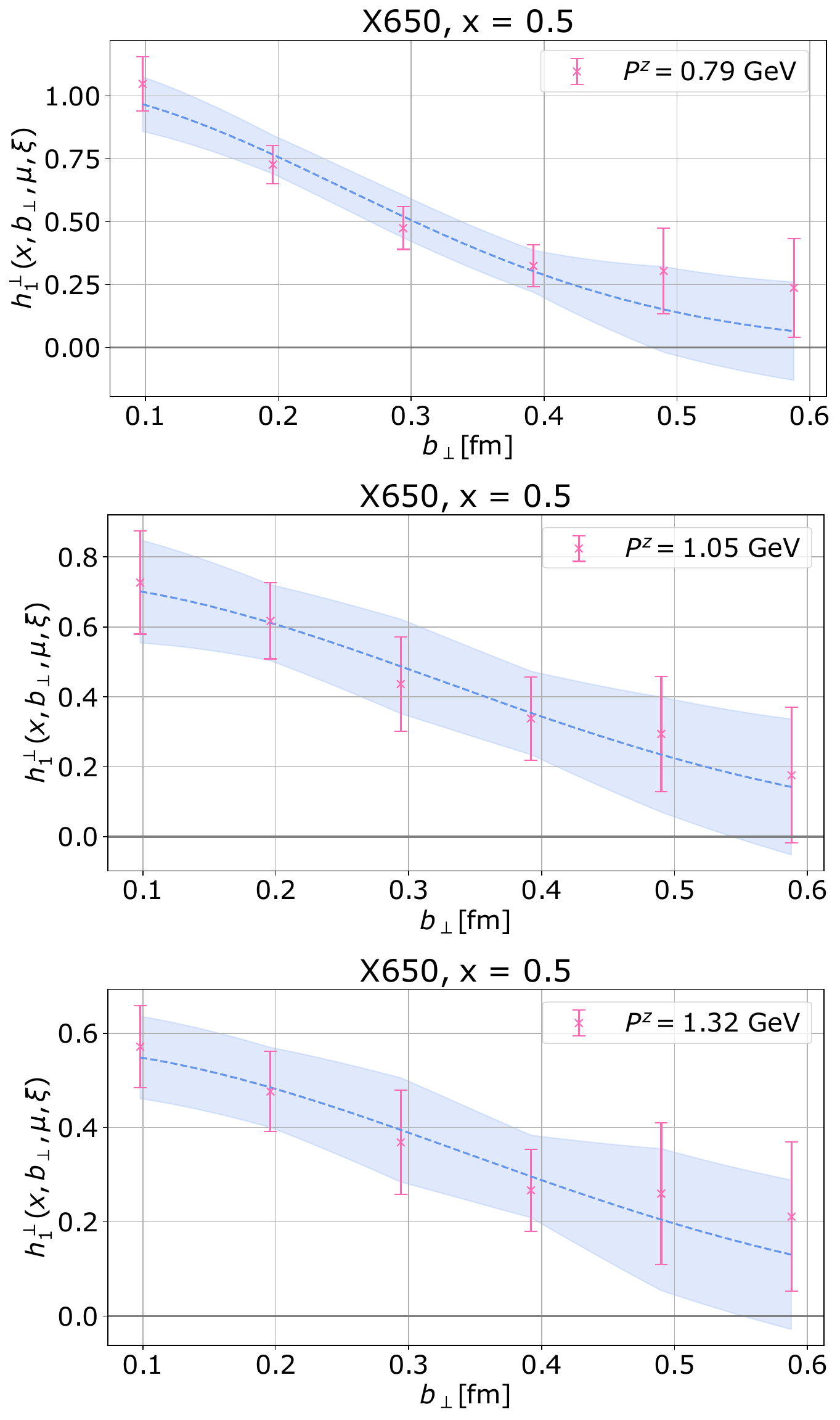}
\caption{Fit of the $b_{\perp}$-dependence of the Boer-Mulders function following Eq.~\eqref{eq:Boer-Mulders_fit_form_b_dep_gauss}. We have taken the X650 result with $x=0.5$ and $P^z=\{0.79,1.05,1.32\}\si{\giga\electronvolt}$ after matching at NNLO as an example.} 
\label{fig:fit_b_dependence}
\end{figure}

An example of fitting the $b_{\perp}$-dependence of the Boer-Mulders function with Eq. \eqref{eq:Boer-Mulders_fit_form_b_dep_gauss} is shown in Fig. ~\ref{fig:fit_b_dependence} for X650 with $x=0.5$ and $P^z=\{0.79,1.05,1.32,1.58\}\si{\giga\electronvolt}$ after matching at NNLO. We choose to display the error obtained from linearly interpolating the bootstrap error of $h_1^{\perp}(x,b_{\perp},\mu,\zeta)$ instead of the fit error to not artificially reduce the error size. As seen from the figure, the Boer-Mulders function decays to zero for $b_{\perp}\approx 0.5-0.6 \si{\femto\meter}$. The resulting fit parameters for X650 with $x=0.5$ are $c_1=\num{1.04 \pm 0.1}$ and $c_2=8.1 (1.6)$ for $P^z=\SI{0.79}{\giga\electronvolt}$, $c_1=\num{0.73 \pm 0.11}$ and $c_2=4.7 (1.8)$ for $P^z=\SI{1.05}{\giga\electronvolt}$, and $c_1=\num{0.57 \pm 0.07}$ and $c_2=4.3 (1.6)$ for $P^z=\SI{1.32}{\giga\electronvolt}$.

In order to compare the results of different ensembles, $h_1^{\perp}(x,b_{\perp},\mu,\zeta)$ is interpolated to $b_{\perp}=\{0.1,0.2,0.3,0.4,0.5,0.6\}\si{\femto\meter}$.

\subsection{Extrapolation to the continuum and infinite momentum}
\label{subsec:extrapolation_p_a}

A combined extrapolation of the Boer-Mulders function to infinite momentum and to the continuum is performed with 
\begin{align}\label{eq:BM-combined-extrap}
h_1^{\perp} (x, b_{\perp}, a, P^z) = &h_{1,0}^{\perp} (x, b_{\perp}) + a^2 f(x, b_{\perp}) \\
&+ a^2 (P^z)^2 h(x, b_{\perp}) + \frac{g(x,b_{\perp},a)}{(P^z)^2},\nonumber
\end{align}
where for simplicity we have suppressed the dependence on renormalization scale. $h_1^{\perp} (x, b_{\perp}, a, P^z)$ on the l.h.s. is the lightcone Boer-Mulders function for different lattice spacings and pion momenta, obtained for specific values of $b_{\perp}$ by fitting the $b_{\perp}$-dependence with Eq. \eqref{eq:Boer-Mulders_fit_form_b_dep_gauss}. Discretization effects are accounted for by the terms $a^2 f(x,b_{\perp})$ and $a^2 (P^z)^2 h(x,b_{\perp})$, while $g(x,b_{\perp},a) / (P^z)^2$ specifies the dependence of the leading power correction on $P^z$. An explicit $a$-dependence is kept in the parameter $g(x,b_{\perp},a)$. The final result for the Boer-Mulders function after extrapolation is given by $h_{1,0}^{\perp} (x, b_{\perp})$. 

\begin{figure*}[thbp]
\includegraphics[width=\textwidth]{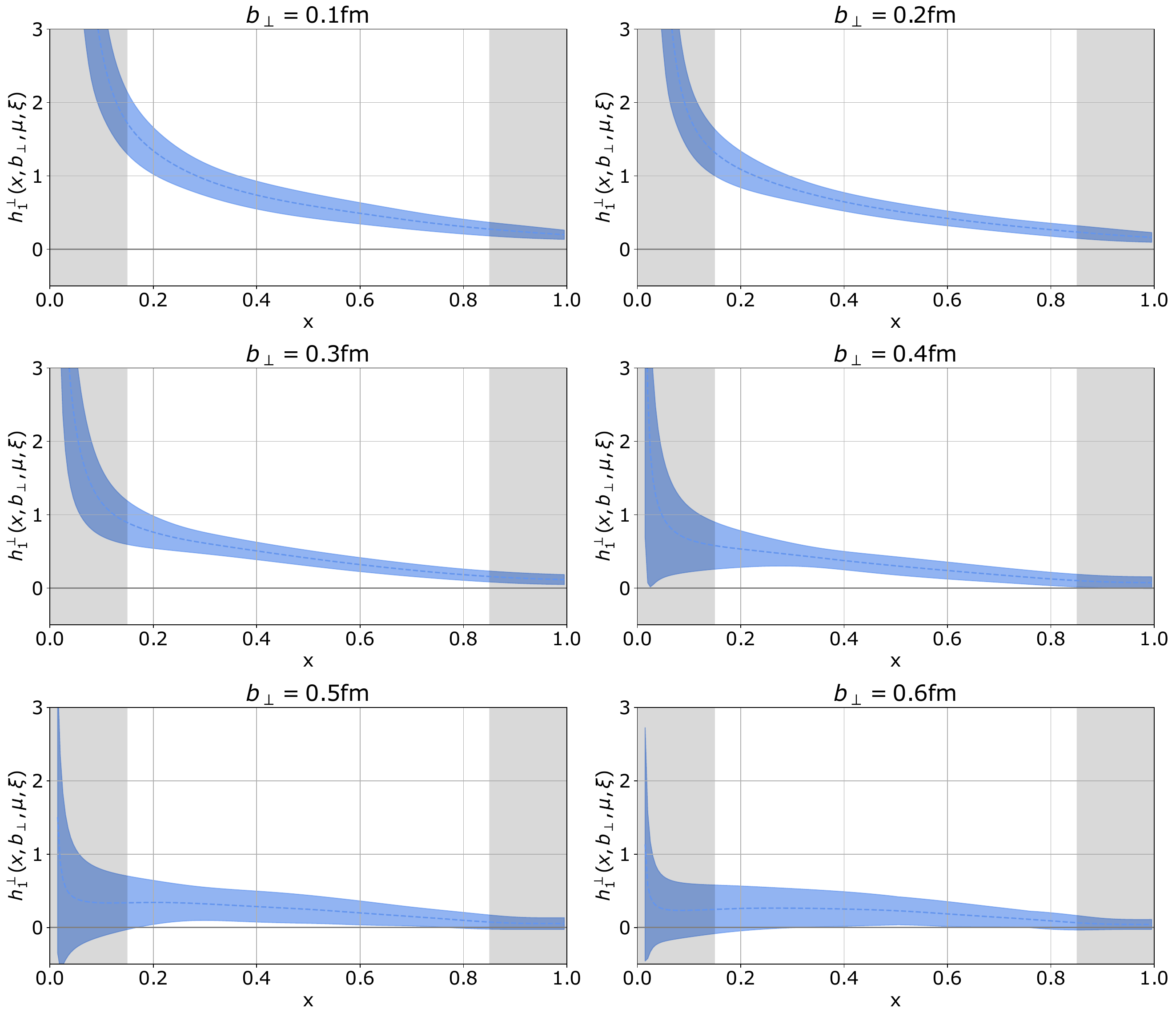}
\caption{Results of the Pion Boer-Mulders function after extrapolation to infinite momentum and to the continuum with Eq.~\eqref{eq:BM-combined-extrap} and NNLO matching kernel. The regions that are shaded in grey indicate the range where the LaMET factorization becomes unreliable due to power corrections. } 
\label{fig:result_extrapolated}
\end{figure*}

Fig. \ref{fig:result_extrapolated} shows the Boer-Mulders function extrapolated to the continuum and infinite momentum limit, with error bands including statistical errors as well as systematic uncertainties from the sources mentioned above. The uncertainty from the extrapolation is determined as the difference between the extrapolated result and the result of N203 at $P^z=\SI{1.61}{\giga\electronvolt}$. It is added in quadrature with the other errors. The regions $x \in [0,0.15]$ and $x \in [0.85,1]$ are shaded in grey to indicate the range where the LaMET factorization becomes unreliable due to power corrections. 

In contrast to the unpolarized quark TMDPDF calculated in Ref.~\cite{LatticePartonCollaborationLPC:2022myp}, a clear decay trend with increasing $b_\perp$ is observed in the pion Boer-Mulders function. The reason could potentially be the following: The pion is much lighter than the nucleon. Therefore, for $P^z\approx 1\sim 2$ GeV the power corrections are already much less important than for the nucleon. On the other hand, according to the analysis in Ref.~\cite{Vladimirov:2020ofp}, the higher-twist contamination for Boer-Mulders function is smaller than for the unpolarized quark TMDPDF.

\section{Summary and outlook}
\label{SEC:summ}

To summarize, we have presented the first lattice calculation of the T-odd Boer-Mulders quark TMDPDF with LaMET. The calculation is done at three lattice spacings with pion mass $~\sim 350$ GeV and momenta up to $1.8$ GeV. The lattice matrix elements are renormalized in a short-distance scheme and extrapolated to the continuum and infinite momentum limit. It is found that the pion Boer-Mulders function decays with increasing $b_{\perp}$ and is compatible with zero for $b_{\perp} \approx 0.5-0.6 \si{\femto\meter}$.

We have also investigated the numerical impact of perturbative matching up to the NNLO and RGR. The perturbative matching mainly affects the small-$x$ region. Using RGR, the Boer-Mulders function cannot be determined for small $x$ due to the breakdown of the resummation. The range of $x$ that can be used after RGR depends on $P^z$, and using the largest momentum $P^z=\SI{1.84}{\giga\electronvolt}$ of our analysis, the regions of $x \in [0,0.15]$ and $x \in [0.85,1]$ are unreliable.

The pion Boer-Mulders function determined in this work will offer valuable guidance for phenomenological analyses of the Boer-Mulders single-spin asymmetry, as well as for future measurements at JLab and EIC.

In the future, our results can be improved in several aspects. First of all, the pion mass dependence can be studied by calculating on ensembles with different pion masses. Secondly, the Collins-Soper kernel and intrinsic soft function are not yet available on the ensembles H102 and N203. Our current analysis is based on an interpolation of the results on X650. They will be calculated in forthcoming publications.

\section*{Acknowledgments}
We thank the CLS Collaboration for sharing the lattices used to perform this study. {The authors gratefully acknowledge the gauss Centre for Supercomputing e.V. (www.gauss-centre.eu) for funding this project by providing computing time on the GCS Supercomputer SuperMUC-NG at Leibniz Supercomputing Centre (www.lrz.de).} The LQCD calculations were performed using the multigrid algorithm~\cite{Babich:2010qb,Osborn:2010mb} and Chroma software
suite~\cite{Edwards:2004sx}. LM and JHZ are supported in part by the National Natural Science Foundation of China under grants No. 12375080, 11975051, and by CUHK-Shenzhen under grant No. UDF01002851.  WW is supported in part by Natural Science Foundation of China under grant No. 12125503 and  12335003. XNX is supported in part by the National Natural Science Foundation of China
under Grant No.~11905296.
YY is supported in part by NSFC grants No. 12293060, 12293062, 12435002 and 12047503, National Key R\&D Program of China No.2024YFE0109800, the Strategic Priority Research Program of Chinese Academy of Sciences, Grant No. XDB34030303 and YSBR-101. QAZ is supported in part by the National Natural Science Foundation of China
under Grant No.~12375069 and the Fundamental Research Funds for the Central Universities.
AS, HTS, PS, WW, YY and JHZ are also supported by a NSFC-DFG joint grant under grant No. 12061131006 and SCHA~458/22. PS is also supported by Strategic Priority Research Program of the Chinese Academy of Sciences under grant number XDB34030301. JH is supported in part by the National Natural Science Foundation of China under grants No. 12205106 and by Guangdong Major Project of Basic and Applied Basic Research No. 2020B0301030008.

\bibliographystyle{apsrev}
\bibliography{ref}

\begin{thebibliography}{75}
\expandafter\ifx\csname natexlab\endcsname\relax\def\natexlab#1{#1}\fi
\expandafter\ifx\csname bibnamefont\endcsname\relax
  \def\bibnamefont#1{#1}\fi
\expandafter\ifx\csname bibfnamefont\endcsname\relax
  \def\bibfnamefont#1{#1}\fi
\expandafter\ifx\csname citenamefont\endcsname\relax
  \def\citenamefont#1{#1}\fi
\expandafter\ifx\csname url\endcsname\relax
  \def\url#1{\texttt{#1}}\fi
\expandafter\ifx\csname urlprefix\endcsname\relax\def\urlprefix{URL }\fi
\providecommand{\bibinfo}[2]{#2}
\providecommand{\eprint}[2][]{\url{#2}}

\bibitem[{\citenamefont{Abdul~Khalek et~al.}(2022)}]{AbdulKhalek:2021gbh}
\bibinfo{author}{\bibfnamefont{R.}~\bibnamefont{Abdul~Khalek}}
  \bibnamefont{et~al.}, \bibinfo{journal}{Nucl. Phys. A}
  \textbf{\bibinfo{volume}{1026}}, \bibinfo{pages}{122447}
  (\bibinfo{year}{2022}), \eprint{2103.05419}.

\bibitem[{\citenamefont{Anderle et~al.}(2021)}]{Anderle:2021wcy}
\bibinfo{author}{\bibfnamefont{D.~P.} \bibnamefont{Anderle}}
  \bibnamefont{et~al.}, \bibinfo{journal}{Front. Phys. (Beijing)}
  \textbf{\bibinfo{volume}{16}}, \bibinfo{pages}{64701} (\bibinfo{year}{2021}),
  \eprint{2102.09222}.

\bibitem[{\citenamefont{Bacchetta et~al.}(2017)\citenamefont{Bacchetta,
  Delcarro, Pisano, Radici, and Signori}}]{Bacchetta:2017gcc}
\bibinfo{author}{\bibfnamefont{A.}~\bibnamefont{Bacchetta}},
  \bibinfo{author}{\bibfnamefont{F.}~\bibnamefont{Delcarro}},
  \bibinfo{author}{\bibfnamefont{C.}~\bibnamefont{Pisano}},
  \bibinfo{author}{\bibfnamefont{M.}~\bibnamefont{Radici}}, \bibnamefont{and}
  \bibinfo{author}{\bibfnamefont{A.}~\bibnamefont{Signori}},
  \bibinfo{journal}{JHEP} \textbf{\bibinfo{volume}{06}}, \bibinfo{pages}{081}
  (\bibinfo{year}{2017}), \bibinfo{note}{[Erratum: JHEP 06, 051 (2019)]},
  \eprint{1703.10157}.

\bibitem[{\citenamefont{Scimemi and Vladimirov}(2018)}]{Scimemi:2017etj}
\bibinfo{author}{\bibfnamefont{I.}~\bibnamefont{Scimemi}} \bibnamefont{and}
  \bibinfo{author}{\bibfnamefont{A.}~\bibnamefont{Vladimirov}},
  \bibinfo{journal}{Eur. Phys. J. C} \textbf{\bibinfo{volume}{78}},
  \bibinfo{pages}{89} (\bibinfo{year}{2018}), \eprint{1706.01473}.

\bibitem[{\citenamefont{Bertone et~al.}(2019)\citenamefont{Bertone, Scimemi,
  and Vladimirov}}]{Bertone:2019nxa}
\bibinfo{author}{\bibfnamefont{V.}~\bibnamefont{Bertone}},
  \bibinfo{author}{\bibfnamefont{I.}~\bibnamefont{Scimemi}}, \bibnamefont{and}
  \bibinfo{author}{\bibfnamefont{A.}~\bibnamefont{Vladimirov}},
  \bibinfo{journal}{JHEP} \textbf{\bibinfo{volume}{06}}, \bibinfo{pages}{028}
  (\bibinfo{year}{2019}), \eprint{1902.08474}.

\bibitem[{\citenamefont{Scimemi and Vladimirov}(2020)}]{Scimemi:2019cmh}
\bibinfo{author}{\bibfnamefont{I.}~\bibnamefont{Scimemi}} \bibnamefont{and}
  \bibinfo{author}{\bibfnamefont{A.}~\bibnamefont{Vladimirov}},
  \bibinfo{journal}{JHEP} \textbf{\bibinfo{volume}{06}}, \bibinfo{pages}{137}
  (\bibinfo{year}{2020}), \eprint{1912.06532}.

\bibitem[{\citenamefont{Bacchetta et~al.}(2020)\citenamefont{Bacchetta,
  Bertone, Bissolotti, Bozzi, Delcarro, Piacenza, and
  Radici}}]{Bacchetta:2019sam}
\bibinfo{author}{\bibfnamefont{A.}~\bibnamefont{Bacchetta}},
  \bibinfo{author}{\bibfnamefont{V.}~\bibnamefont{Bertone}},
  \bibinfo{author}{\bibfnamefont{C.}~\bibnamefont{Bissolotti}},
  \bibinfo{author}{\bibfnamefont{G.}~\bibnamefont{Bozzi}},
  \bibinfo{author}{\bibfnamefont{F.}~\bibnamefont{Delcarro}},
  \bibinfo{author}{\bibfnamefont{F.}~\bibnamefont{Piacenza}}, \bibnamefont{and}
  \bibinfo{author}{\bibfnamefont{M.}~\bibnamefont{Radici}},
  \bibinfo{journal}{JHEP} \textbf{\bibinfo{volume}{07}}, \bibinfo{pages}{117}
  (\bibinfo{year}{2020}), \eprint{1912.07550}.

\bibitem[{\citenamefont{Bacchetta et~al.}(2022)\citenamefont{Bacchetta,
  Delcarro, Pisano, and Radici}}]{Bacchetta:2020gko}
\bibinfo{author}{\bibfnamefont{A.}~\bibnamefont{Bacchetta}},
  \bibinfo{author}{\bibfnamefont{F.}~\bibnamefont{Delcarro}},
  \bibinfo{author}{\bibfnamefont{C.}~\bibnamefont{Pisano}}, \bibnamefont{and}
  \bibinfo{author}{\bibfnamefont{M.}~\bibnamefont{Radici}},
  \bibinfo{journal}{Phys. Lett. B} \textbf{\bibinfo{volume}{827}},
  \bibinfo{pages}{136961} (\bibinfo{year}{2022}), \eprint{2004.14278}.

\bibitem[{\citenamefont{Bury et~al.}(2021)\citenamefont{Bury, Prokudin, and
  Vladimirov}}]{Bury:2021sue}
\bibinfo{author}{\bibfnamefont{M.}~\bibnamefont{Bury}},
  \bibinfo{author}{\bibfnamefont{A.}~\bibnamefont{Prokudin}}, \bibnamefont{and}
  \bibinfo{author}{\bibfnamefont{A.}~\bibnamefont{Vladimirov}},
  \bibinfo{journal}{JHEP} \textbf{\bibinfo{volume}{05}}, \bibinfo{pages}{151}
  (\bibinfo{year}{2021}), \eprint{2103.03270}.

\bibitem[{\citenamefont{Echevarria et~al.}(2021)\citenamefont{Echevarria, Kang,
  and Terry}}]{Echevarria:2020hpy}
\bibinfo{author}{\bibfnamefont{M.~G.} \bibnamefont{Echevarria}},
  \bibinfo{author}{\bibfnamefont{Z.-B.} \bibnamefont{Kang}}, \bibnamefont{and}
  \bibinfo{author}{\bibfnamefont{J.}~\bibnamefont{Terry}},
  \bibinfo{journal}{JHEP} \textbf{\bibinfo{volume}{01}}, \bibinfo{pages}{126}
  (\bibinfo{year}{2021}), \eprint{2009.10710}.

\bibitem[{\citenamefont{Hägler et~al.}(2009)\citenamefont{Hägler, Musch,
  Negele, and Schäfer}}]{Hagler:2009mb}
\bibinfo{author}{\bibfnamefont{P.}~\bibnamefont{Hägler}},
  \bibinfo{author}{\bibfnamefont{B.~U.} \bibnamefont{Musch}},
  \bibinfo{author}{\bibfnamefont{J.~W.} \bibnamefont{Negele}},
  \bibnamefont{and} \bibinfo{author}{\bibfnamefont{A.}~\bibnamefont{Schäfer}},
  \bibinfo{journal}{EPL} \textbf{\bibinfo{volume}{88}}, \bibinfo{pages}{61001}
  (\bibinfo{year}{2009}), \eprint{0908.1283}.

\bibitem[{\citenamefont{Musch et~al.}(2011)\citenamefont{Musch, Hägler,
  Negele, and Schäfer}}]{Musch:2010ka}
\bibinfo{author}{\bibfnamefont{B.~U.} \bibnamefont{Musch}},
  \bibinfo{author}{\bibfnamefont{P.}~\bibnamefont{Hägler}},
  \bibinfo{author}{\bibfnamefont{J.~W.} \bibnamefont{Negele}},
  \bibnamefont{and} \bibinfo{author}{\bibfnamefont{A.}~\bibnamefont{Schäfer}},
  \bibinfo{journal}{Phys. Rev. D} \textbf{\bibinfo{volume}{83}},
  \bibinfo{pages}{094507} (\bibinfo{year}{2011}), \eprint{1011.1213}.

\bibitem[{\citenamefont{Musch et~al.}(2012)\citenamefont{Musch, Hägler,
  Engelhardt, Negele, and Schäfer}}]{Musch:2011er}
\bibinfo{author}{\bibfnamefont{B.~U.} \bibnamefont{Musch}},
  \bibinfo{author}{\bibfnamefont{P.}~\bibnamefont{Hägler}},
  \bibinfo{author}{\bibfnamefont{M.}~\bibnamefont{Engelhardt}},
  \bibinfo{author}{\bibfnamefont{J.~W.} \bibnamefont{Negele}},
  \bibnamefont{and} \bibinfo{author}{\bibfnamefont{A.}~\bibnamefont{Schäfer}},
  \bibinfo{journal}{Phys. Rev. D} \textbf{\bibinfo{volume}{85}},
  \bibinfo{pages}{094510} (\bibinfo{year}{2012}), \eprint{1111.4249}.

\bibitem[{\citenamefont{Engelhardt et~al.}(2016)\citenamefont{Engelhardt,
  H\"agler, Musch, Negele, and Sch\"afer}}]{Engelhardt:2015xja}
\bibinfo{author}{\bibfnamefont{M.}~\bibnamefont{Engelhardt}},
  \bibinfo{author}{\bibfnamefont{P.}~\bibnamefont{H\"agler}},
  \bibinfo{author}{\bibfnamefont{B.}~\bibnamefont{Musch}},
  \bibinfo{author}{\bibfnamefont{J.}~\bibnamefont{Negele}}, \bibnamefont{and}
  \bibinfo{author}{\bibfnamefont{A.}~\bibnamefont{Sch\"afer}},
  \bibinfo{journal}{Phys. Rev. D} \textbf{\bibinfo{volume}{93}},
  \bibinfo{pages}{054501} (\bibinfo{year}{2016}), \eprint{1506.07826}.

\bibitem[{\citenamefont{Yoon et~al.}(2017)\citenamefont{Yoon, Engelhardt,
  Gupta, Bhattacharya, Green, Musch, Negele, Pochinsky, Sch\"afer, and
  Syritsyn}}]{Yoon:2017qzo}
\bibinfo{author}{\bibfnamefont{B.}~\bibnamefont{Yoon}},
  \bibinfo{author}{\bibfnamefont{M.}~\bibnamefont{Engelhardt}},
  \bibinfo{author}{\bibfnamefont{R.}~\bibnamefont{Gupta}},
  \bibinfo{author}{\bibfnamefont{T.}~\bibnamefont{Bhattacharya}},
  \bibinfo{author}{\bibfnamefont{J.~R.} \bibnamefont{Green}},
  \bibinfo{author}{\bibfnamefont{B.~U.} \bibnamefont{Musch}},
  \bibinfo{author}{\bibfnamefont{J.~W.} \bibnamefont{Negele}},
  \bibinfo{author}{\bibfnamefont{A.~V.} \bibnamefont{Pochinsky}},
  \bibinfo{author}{\bibfnamefont{A.}~\bibnamefont{Sch\"afer}},
  \bibnamefont{and} \bibinfo{author}{\bibfnamefont{S.~N.}
  \bibnamefont{Syritsyn}}, \bibinfo{journal}{Phys. Rev. D}
  \textbf{\bibinfo{volume}{96}}, \bibinfo{pages}{094508}
  (\bibinfo{year}{2017}), \eprint{1706.03406}.

\bibitem[{\citenamefont{Ji}(2013)}]{Ji:2013dva}
\bibinfo{author}{\bibfnamefont{X.}~\bibnamefont{Ji}}, \bibinfo{journal}{Phys.
  Rev. Lett.} \textbf{\bibinfo{volume}{110}}, \bibinfo{pages}{262002}
  (\bibinfo{year}{2013}), \eprint{1305.1539}.

\bibitem[{\citenamefont{Ji}(2014)}]{Ji:2014gla}
\bibinfo{author}{\bibfnamefont{X.}~\bibnamefont{Ji}}, \bibinfo{journal}{Sci.
  China Phys. Mech. Astron.} \textbf{\bibinfo{volume}{57}},
  \bibinfo{pages}{1407} (\bibinfo{year}{2014}), \eprint{1404.6680}.

\bibitem[{\citenamefont{Ji et~al.}(2021{\natexlab{a}})\citenamefont{Ji, Liu,
  Liu, Zhang, and Zhao}}]{Ji:2020ect}
\bibinfo{author}{\bibfnamefont{X.}~\bibnamefont{Ji}},
  \bibinfo{author}{\bibfnamefont{Y.-S.} \bibnamefont{Liu}},
  \bibinfo{author}{\bibfnamefont{Y.}~\bibnamefont{Liu}},
  \bibinfo{author}{\bibfnamefont{J.-H.} \bibnamefont{Zhang}}, \bibnamefont{and}
  \bibinfo{author}{\bibfnamefont{Y.}~\bibnamefont{Zhao}},
  \bibinfo{journal}{Rev. Mod. Phys.} \textbf{\bibinfo{volume}{93}},
  \bibinfo{pages}{035005} (\bibinfo{year}{2021}{\natexlab{a}}),
  \eprint{2004.03543}.

\bibitem[{\citenamefont{Ji et~al.}(2015)\citenamefont{Ji, Sun, Xiong, and
  Yuan}}]{Ji:2014hxa}
\bibinfo{author}{\bibfnamefont{X.}~\bibnamefont{Ji}},
  \bibinfo{author}{\bibfnamefont{P.}~\bibnamefont{Sun}},
  \bibinfo{author}{\bibfnamefont{X.}~\bibnamefont{Xiong}}, \bibnamefont{and}
  \bibinfo{author}{\bibfnamefont{F.}~\bibnamefont{Yuan}},
  \bibinfo{journal}{Phys. Rev.} \textbf{\bibinfo{volume}{D91}},
  \bibinfo{pages}{074009} (\bibinfo{year}{2015}), \eprint{1405.7640}.

\bibitem[{\citenamefont{Ji et~al.}(2019)\citenamefont{Ji, Jin, Yuan, Zhang, and
  Zhao}}]{Ji:2018hvs}
\bibinfo{author}{\bibfnamefont{X.}~\bibnamefont{Ji}},
  \bibinfo{author}{\bibfnamefont{L.-C.} \bibnamefont{Jin}},
  \bibinfo{author}{\bibfnamefont{F.}~\bibnamefont{Yuan}},
  \bibinfo{author}{\bibfnamefont{J.-H.} \bibnamefont{Zhang}}, \bibnamefont{and}
  \bibinfo{author}{\bibfnamefont{Y.}~\bibnamefont{Zhao}},
  \bibinfo{journal}{Phys. Rev. D} \textbf{\bibinfo{volume}{99}},
  \bibinfo{pages}{114006} (\bibinfo{year}{2019}), \eprint{1801.05930}.

\bibitem[{\citenamefont{Ji et~al.}(2020{\natexlab{a}})\citenamefont{Ji, Liu,
  and Liu}}]{Ji:2019sxk}
\bibinfo{author}{\bibfnamefont{X.}~\bibnamefont{Ji}},
  \bibinfo{author}{\bibfnamefont{Y.}~\bibnamefont{Liu}}, \bibnamefont{and}
  \bibinfo{author}{\bibfnamefont{Y.-S.} \bibnamefont{Liu}},
  \bibinfo{journal}{Nucl. Phys. B} \textbf{\bibinfo{volume}{955}},
  \bibinfo{pages}{115054} (\bibinfo{year}{2020}{\natexlab{a}}),
  \eprint{1910.11415}.

\bibitem[{\citenamefont{Ji et~al.}(2020{\natexlab{b}})\citenamefont{Ji, Liu,
  and Liu}}]{Ji:2019ewn}
\bibinfo{author}{\bibfnamefont{X.}~\bibnamefont{Ji}},
  \bibinfo{author}{\bibfnamefont{Y.}~\bibnamefont{Liu}}, \bibnamefont{and}
  \bibinfo{author}{\bibfnamefont{Y.-S.} \bibnamefont{Liu}},
  \bibinfo{journal}{Phys. Lett. B} \textbf{\bibinfo{volume}{811}},
  \bibinfo{pages}{135946} (\bibinfo{year}{2020}{\natexlab{b}}),
  \eprint{1911.03840}.

\bibitem[{\citenamefont{Ji et~al.}(2021{\natexlab{b}})\citenamefont{Ji, Liu,
  Sch\"afer, and Yuan}}]{Ji:2020jeb}
\bibinfo{author}{\bibfnamefont{X.}~\bibnamefont{Ji}},
  \bibinfo{author}{\bibfnamefont{Y.}~\bibnamefont{Liu}},
  \bibinfo{author}{\bibfnamefont{A.}~\bibnamefont{Sch\"afer}},
  \bibnamefont{and} \bibinfo{author}{\bibfnamefont{F.}~\bibnamefont{Yuan}},
  \bibinfo{journal}{Phys. Rev. D} \textbf{\bibinfo{volume}{103}},
  \bibinfo{pages}{074005} (\bibinfo{year}{2021}{\natexlab{b}}),
  \eprint{2011.13397}.

\bibitem[{\citenamefont{Zhang et~al.}(2020{\natexlab{a}})}]{Zhang:2020dbb}
\bibinfo{author}{\bibfnamefont{Q.-A.} \bibnamefont{Zhang}} \bibnamefont{et~al.}
  (\bibinfo{collaboration}{Lattice Parton}), \bibinfo{journal}{Phys. Rev.
  Lett.} \textbf{\bibinfo{volume}{125}}, \bibinfo{pages}{192001}
  (\bibinfo{year}{2020}{\natexlab{a}}), \eprint{2005.14572}.

\bibitem[{\citenamefont{Ebert et~al.}(2019{\natexlab{a}})\citenamefont{Ebert,
  Stewart, and Zhao}}]{Ebert:2018gzl}
\bibinfo{author}{\bibfnamefont{M.~A.} \bibnamefont{Ebert}},
  \bibinfo{author}{\bibfnamefont{I.~W.} \bibnamefont{Stewart}},
  \bibnamefont{and} \bibinfo{author}{\bibfnamefont{Y.}~\bibnamefont{Zhao}},
  \bibinfo{journal}{Phys. Rev. D} \textbf{\bibinfo{volume}{99}},
  \bibinfo{pages}{034505} (\bibinfo{year}{2019}{\natexlab{a}}),
  \eprint{1811.00026}.

\bibitem[{\citenamefont{Ebert et~al.}(2019{\natexlab{b}})\citenamefont{Ebert,
  Stewart, and Zhao}}]{Ebert:2019okf}
\bibinfo{author}{\bibfnamefont{M.~A.} \bibnamefont{Ebert}},
  \bibinfo{author}{\bibfnamefont{I.~W.} \bibnamefont{Stewart}},
  \bibnamefont{and} \bibinfo{author}{\bibfnamefont{Y.}~\bibnamefont{Zhao}},
  \bibinfo{journal}{JHEP} \textbf{\bibinfo{volume}{09}}, \bibinfo{pages}{037}
  (\bibinfo{year}{2019}{\natexlab{b}}), \eprint{1901.03685}.

\bibitem[{\citenamefont{Ebert et~al.}(2020{\natexlab{a}})\citenamefont{Ebert,
  Stewart, and Zhao}}]{Ebert:2019tvc}
\bibinfo{author}{\bibfnamefont{M.~A.} \bibnamefont{Ebert}},
  \bibinfo{author}{\bibfnamefont{I.~W.} \bibnamefont{Stewart}},
  \bibnamefont{and} \bibinfo{author}{\bibfnamefont{Y.}~\bibnamefont{Zhao}},
  \bibinfo{journal}{JHEP} \textbf{\bibinfo{volume}{03}}, \bibinfo{pages}{099}
  (\bibinfo{year}{2020}{\natexlab{a}}), \eprint{1910.08569}.

\bibitem[{\citenamefont{Shanahan
  et~al.}(2020{\natexlab{a}})\citenamefont{Shanahan, Wagman, and
  Zhao}}]{Shanahan:2019zcq}
\bibinfo{author}{\bibfnamefont{P.}~\bibnamefont{Shanahan}},
  \bibinfo{author}{\bibfnamefont{M.~L.} \bibnamefont{Wagman}},
  \bibnamefont{and} \bibinfo{author}{\bibfnamefont{Y.}~\bibnamefont{Zhao}},
  \bibinfo{journal}{Phys. Rev. D} \textbf{\bibinfo{volume}{101}},
  \bibinfo{pages}{074505} (\bibinfo{year}{2020}{\natexlab{a}}),
  \eprint{1911.00800}.

\bibitem[{\citenamefont{Ebert et~al.}(2020{\natexlab{b}})\citenamefont{Ebert,
  Schindler, Stewart, and Zhao}}]{Ebert:2020gxr}
\bibinfo{author}{\bibfnamefont{M.~A.} \bibnamefont{Ebert}},
  \bibinfo{author}{\bibfnamefont{S.~T.} \bibnamefont{Schindler}},
  \bibinfo{author}{\bibfnamefont{I.~W.} \bibnamefont{Stewart}},
  \bibnamefont{and} \bibinfo{author}{\bibfnamefont{Y.}~\bibnamefont{Zhao}},
  \bibinfo{journal}{JHEP} \textbf{\bibinfo{volume}{09}}, \bibinfo{pages}{099}
  (\bibinfo{year}{2020}{\natexlab{b}}), \eprint{2004.14831}.

\bibitem[{\citenamefont{Vladimirov and Sch\"afer}(2020)}]{Vladimirov:2020ofp}
\bibinfo{author}{\bibfnamefont{A.~A.} \bibnamefont{Vladimirov}}
  \bibnamefont{and}
  \bibinfo{author}{\bibfnamefont{A.}~\bibnamefont{Sch\"afer}},
  \bibinfo{journal}{Phys. Rev. D} \textbf{\bibinfo{volume}{101}},
  \bibinfo{pages}{074517} (\bibinfo{year}{2020}), \eprint{2002.07527}.

\bibitem[{\citenamefont{Ji et~al.}(2021{\natexlab{c}})\citenamefont{Ji, Zhang,
  Zhao, and Zhu}}]{Ji:2021uvr}
\bibinfo{author}{\bibfnamefont{Y.}~\bibnamefont{Ji}},
  \bibinfo{author}{\bibfnamefont{J.-H.} \bibnamefont{Zhang}},
  \bibinfo{author}{\bibfnamefont{S.}~\bibnamefont{Zhao}}, \bibnamefont{and}
  \bibinfo{author}{\bibfnamefont{R.}~\bibnamefont{Zhu}},
  \bibinfo{journal}{Phys. Rev. D} \textbf{\bibinfo{volume}{104}},
  \bibinfo{pages}{094510} (\bibinfo{year}{2021}{\natexlab{c}}),
  \eprint{2104.13345}.

\bibitem[{\citenamefont{Ebert et~al.}(2022)\citenamefont{Ebert, Schindler,
  Stewart, and Zhao}}]{Ebert:2022fmh}
\bibinfo{author}{\bibfnamefont{M.~A.} \bibnamefont{Ebert}},
  \bibinfo{author}{\bibfnamefont{S.~T.} \bibnamefont{Schindler}},
  \bibinfo{author}{\bibfnamefont{I.~W.} \bibnamefont{Stewart}},
  \bibnamefont{and} \bibinfo{author}{\bibfnamefont{Y.}~\bibnamefont{Zhao}},
  \bibinfo{journal}{JHEP} \textbf{\bibinfo{volume}{04}}, \bibinfo{pages}{178}
  (\bibinfo{year}{2022}), \eprint{2201.08401}.

\bibitem[{\citenamefont{Schindler et~al.}(2022)\citenamefont{Schindler,
  Stewart, and Zhao}}]{Schindler:2022eva}
\bibinfo{author}{\bibfnamefont{S.~T.} \bibnamefont{Schindler}},
  \bibinfo{author}{\bibfnamefont{I.~W.} \bibnamefont{Stewart}},
  \bibnamefont{and} \bibinfo{author}{\bibfnamefont{Y.}~\bibnamefont{Zhao}},
  \bibinfo{journal}{JHEP} \textbf{\bibinfo{volume}{08}}, \bibinfo{pages}{084}
  (\bibinfo{year}{2022}), \eprint{2205.12369}.

\bibitem[{\citenamefont{Zhu et~al.}(2023)\citenamefont{Zhu, Ji, Zhang, and
  Zhao}}]{Zhu:2022bja}
\bibinfo{author}{\bibfnamefont{R.}~\bibnamefont{Zhu}},
  \bibinfo{author}{\bibfnamefont{Y.}~\bibnamefont{Ji}},
  \bibinfo{author}{\bibfnamefont{J.-H.} \bibnamefont{Zhang}}, \bibnamefont{and}
  \bibinfo{author}{\bibfnamefont{S.}~\bibnamefont{Zhao}},
  \bibinfo{journal}{JHEP} \textbf{\bibinfo{volume}{02}}, \bibinfo{pages}{114}
  (\bibinfo{year}{2023}), \eprint{2209.05443}.

\bibitem[{\citenamefont{Zhao}(2023)}]{Zhao:2023ptv}
\bibinfo{author}{\bibfnamefont{Y.}~\bibnamefont{Zhao}} (\bibinfo{year}{2023}),
  \eprint{2311.01391}.

\bibitem[{\citenamefont{Han et~al.}(2024{\natexlab{a}})\citenamefont{Han, Hua,
  Ji, L\"u, Wang, Xu, Zhang, and Zhao}}]{Han:2024min}
\bibinfo{author}{\bibfnamefont{X.-Y.} \bibnamefont{Han}},
  \bibinfo{author}{\bibfnamefont{J.}~\bibnamefont{Hua}},
  \bibinfo{author}{\bibfnamefont{X.}~\bibnamefont{Ji}},
  \bibinfo{author}{\bibfnamefont{C.-D.} \bibnamefont{L\"u}},
  \bibinfo{author}{\bibfnamefont{W.}~\bibnamefont{Wang}},
  \bibinfo{author}{\bibfnamefont{J.}~\bibnamefont{Xu}},
  \bibinfo{author}{\bibfnamefont{Q.-A.} \bibnamefont{Zhang}}, \bibnamefont{and}
  \bibinfo{author}{\bibfnamefont{S.}~\bibnamefont{Zhao}}
  (\bibinfo{year}{2024}{\natexlab{a}}), \eprint{2403.17492}.

\bibitem[{\citenamefont{Shanahan
  et~al.}(2020{\natexlab{b}})\citenamefont{Shanahan, Wagman, and
  Zhao}}]{Shanahan:2020zxr}
\bibinfo{author}{\bibfnamefont{P.}~\bibnamefont{Shanahan}},
  \bibinfo{author}{\bibfnamefont{M.}~\bibnamefont{Wagman}}, \bibnamefont{and}
  \bibinfo{author}{\bibfnamefont{Y.}~\bibnamefont{Zhao}},
  \bibinfo{journal}{Phys. Rev. D} \textbf{\bibinfo{volume}{102}},
  \bibinfo{pages}{014511} (\bibinfo{year}{2020}{\natexlab{b}}),
  \eprint{2003.06063}.

\bibitem[{\citenamefont{Zhang
  et~al.}(2020{\natexlab{b}})}]{LatticeParton:2020uhz}
\bibinfo{author}{\bibfnamefont{Q.-A.} \bibnamefont{Zhang}} \bibnamefont{et~al.}
  (\bibinfo{collaboration}{Lattice Parton}), \bibinfo{journal}{Phys. Rev.
  Lett.} \textbf{\bibinfo{volume}{125}}, \bibinfo{pages}{192001}
  (\bibinfo{year}{2020}{\natexlab{b}}), \eprint{2005.14572}.

\bibitem[{\citenamefont{Schlemmer et~al.}(2021)\citenamefont{Schlemmer,
  Vladimirov, Zimmermann, Engelhardt, and Sch\"afer}}]{Schlemmer:2021aij}
\bibinfo{author}{\bibfnamefont{M.}~\bibnamefont{Schlemmer}},
  \bibinfo{author}{\bibfnamefont{A.}~\bibnamefont{Vladimirov}},
  \bibinfo{author}{\bibfnamefont{C.}~\bibnamefont{Zimmermann}},
  \bibinfo{author}{\bibfnamefont{M.}~\bibnamefont{Engelhardt}},
  \bibnamefont{and}
  \bibinfo{author}{\bibfnamefont{A.}~\bibnamefont{Sch\"afer}},
  \bibinfo{journal}{JHEP} \textbf{\bibinfo{volume}{08}}, \bibinfo{pages}{004}
  (\bibinfo{year}{2021}), \eprint{2103.16991}.

\bibitem[{\citenamefont{Shanahan et~al.}(2021)\citenamefont{Shanahan, Wagman,
  and Zhao}}]{Shanahan:2021tst}
\bibinfo{author}{\bibfnamefont{P.}~\bibnamefont{Shanahan}},
  \bibinfo{author}{\bibfnamefont{M.}~\bibnamefont{Wagman}}, \bibnamefont{and}
  \bibinfo{author}{\bibfnamefont{Y.}~\bibnamefont{Zhao}},
  \bibinfo{journal}{Phys. Rev. D} \textbf{\bibinfo{volume}{104}},
  \bibinfo{pages}{114502} (\bibinfo{year}{2021}), \eprint{2107.11930}.

\bibitem[{\citenamefont{Chu et~al.}(2022)}]{LatticePartonLPC:2022eev}
\bibinfo{author}{\bibfnamefont{M.-H.} \bibnamefont{Chu}} \bibnamefont{et~al.}
  (\bibinfo{collaboration}{Lattice Parton (LPC)}), \bibinfo{journal}{Phys. Rev.
  D} \textbf{\bibinfo{volume}{106}}, \bibinfo{pages}{034509}
  (\bibinfo{year}{2022}), \eprint{2204.00200}.

\bibitem[{\citenamefont{He et~al.}(2024)\citenamefont{He, Chu, Hua, Ji,
  Sch\"afer, Su, Wang, Yang, Zhang, and
  Zhang}}]{LatticePartonCollaborationLPC:2022myp}
\bibinfo{author}{\bibfnamefont{J.-C.} \bibnamefont{He}},
  \bibinfo{author}{\bibfnamefont{M.-H.} \bibnamefont{Chu}},
  \bibinfo{author}{\bibfnamefont{J.}~\bibnamefont{Hua}},
  \bibinfo{author}{\bibfnamefont{X.}~\bibnamefont{Ji}},
  \bibinfo{author}{\bibfnamefont{A.}~\bibnamefont{Sch\"afer}},
  \bibinfo{author}{\bibfnamefont{Y.}~\bibnamefont{Su}},
  \bibinfo{author}{\bibfnamefont{W.}~\bibnamefont{Wang}},
  \bibinfo{author}{\bibfnamefont{Y.-B.} \bibnamefont{Yang}},
  \bibinfo{author}{\bibfnamefont{J.-H.} \bibnamefont{Zhang}}, \bibnamefont{and}
  \bibinfo{author}{\bibfnamefont{Q.-A.} \bibnamefont{Zhang}}
  (\bibinfo{collaboration}{Lattice Parton Collaboration (LPC)}),
  \bibinfo{journal}{Phys. Rev. D} \textbf{\bibinfo{volume}{109}},
  \bibinfo{pages}{114513} (\bibinfo{year}{2024}), \eprint{2211.02340}.

\bibitem[{\citenamefont{Chu et~al.}(2024)}]{LatticeParton:2023xdl}
\bibinfo{author}{\bibfnamefont{M.-H.} \bibnamefont{Chu}} \bibnamefont{et~al.}
  (\bibinfo{collaboration}{Lattice Parton}), \bibinfo{journal}{Phys. Rev. D}
  \textbf{\bibinfo{volume}{109}}, \bibinfo{pages}{L091503}
  (\bibinfo{year}{2024}), \eprint{2302.09961}.

\bibitem[{\citenamefont{Avkhadiev et~al.}(2023)\citenamefont{Avkhadiev,
  Shanahan, Wagman, and Zhao}}]{Avkhadiev:2023poz}
\bibinfo{author}{\bibfnamefont{A.}~\bibnamefont{Avkhadiev}},
  \bibinfo{author}{\bibfnamefont{P.~E.} \bibnamefont{Shanahan}},
  \bibinfo{author}{\bibfnamefont{M.~L.} \bibnamefont{Wagman}},
  \bibnamefont{and} \bibinfo{author}{\bibfnamefont{Y.}~\bibnamefont{Zhao}},
  \bibinfo{journal}{Phys. Rev. D} \textbf{\bibinfo{volume}{108}},
  \bibinfo{pages}{114505} (\bibinfo{year}{2023}), \eprint{2307.12359}.

\bibitem[{\citenamefont{Avkhadiev et~al.}(2024)\citenamefont{Avkhadiev,
  Shanahan, Wagman, and Zhao}}]{Avkhadiev:2024mgd}
\bibinfo{author}{\bibfnamefont{A.}~\bibnamefont{Avkhadiev}},
  \bibinfo{author}{\bibfnamefont{P.~E.} \bibnamefont{Shanahan}},
  \bibinfo{author}{\bibfnamefont{M.~L.} \bibnamefont{Wagman}},
  \bibnamefont{and} \bibinfo{author}{\bibfnamefont{Y.}~\bibnamefont{Zhao}},
  \bibinfo{journal}{Phys. Rev. Lett.} \textbf{\bibinfo{volume}{132}},
  \bibinfo{pages}{231901} (\bibinfo{year}{2024}), \eprint{2402.06725}.

\bibitem[{\citenamefont{Liu et~al.}(2020)}]{LatticeParton:2018gjr}
\bibinfo{author}{\bibfnamefont{Y.-S.} \bibnamefont{Liu}} \bibnamefont{et~al.}
  (\bibinfo{collaboration}{Lattice Parton}), \bibinfo{journal}{Phys. Rev. D}
  \textbf{\bibinfo{volume}{101}}, \bibinfo{pages}{034020}
  (\bibinfo{year}{2020}), \eprint{1807.06566}.

\bibitem[{\citenamefont{Han et~al.}(2024{\natexlab{b}})}]{Han:2024yun}
\bibinfo{author}{\bibfnamefont{X.-Y.} \bibnamefont{Han}} \bibnamefont{et~al.}
  (\bibinfo{year}{2024}{\natexlab{b}}), \eprint{2410.18654}.

\bibitem[{\citenamefont{del R\'\i{}o and Vladimirov}(2023)}]{delRio:2023pse}
\bibinfo{author}{\bibfnamefont{O.}~\bibnamefont{del R\'\i{}o}}
  \bibnamefont{and}
  \bibinfo{author}{\bibfnamefont{A.}~\bibnamefont{Vladimirov}},
  \bibinfo{journal}{Phys. Rev. D} \textbf{\bibinfo{volume}{108}},
  \bibinfo{pages}{114009} (\bibinfo{year}{2023}), \eprint{2304.14440}.

\bibitem[{\citenamefont{Ji et~al.}(2023)\citenamefont{Ji, Liu, and
  Su}}]{Ji:2023pba}
\bibinfo{author}{\bibfnamefont{X.}~\bibnamefont{Ji}},
  \bibinfo{author}{\bibfnamefont{Y.}~\bibnamefont{Liu}}, \bibnamefont{and}
  \bibinfo{author}{\bibfnamefont{Y.}~\bibnamefont{Su}}, \bibinfo{journal}{JHEP}
  \textbf{\bibinfo{volume}{08}}, \bibinfo{pages}{037} (\bibinfo{year}{2023}),
  \eprint{2305.04416}.

\bibitem[{\citenamefont{Zhang et~al.}(2022)\citenamefont{Zhang, Ji, Yang, Yao,
  and Zhang}}]{Zhang:2022xuw}
\bibinfo{author}{\bibfnamefont{K.}~\bibnamefont{Zhang}},
  \bibinfo{author}{\bibfnamefont{X.}~\bibnamefont{Ji}},
  \bibinfo{author}{\bibfnamefont{Y.-B.} \bibnamefont{Yang}},
  \bibinfo{author}{\bibfnamefont{F.}~\bibnamefont{Yao}}, \bibnamefont{and}
  \bibinfo{author}{\bibfnamefont{J.-H.} \bibnamefont{Zhang}}
  (\bibinfo{collaboration}{Lattice Parton (LPC)}), \bibinfo{journal}{Phys. Rev.
  Lett.} \textbf{\bibinfo{volume}{129}}, \bibinfo{pages}{082002}
  (\bibinfo{year}{2022}), \eprint{2205.13402}.

\bibitem[{\citenamefont{Chu et~al.}(2023)}]{LatticePartonLPC:2023pdv}
\bibinfo{author}{\bibfnamefont{M.-H.} \bibnamefont{Chu}} \bibnamefont{et~al.}
  (\bibinfo{collaboration}{Lattice Parton (LPC)}), \bibinfo{journal}{JHEP}
  \textbf{\bibinfo{volume}{08}}, \bibinfo{pages}{172} (\bibinfo{year}{2023}),
  \eprint{2306.06488}.

\bibitem[{\citenamefont{Ji and Musolf}(1991)}]{Ji:1991pr}
\bibinfo{author}{\bibfnamefont{X.-D.} \bibnamefont{Ji}} \bibnamefont{and}
  \bibinfo{author}{\bibfnamefont{M.~J.} \bibnamefont{Musolf}},
  \bibinfo{journal}{Phys. Lett. B} \textbf{\bibinfo{volume}{257}},
  \bibinfo{pages}{409} (\bibinfo{year}{1991}).

\bibitem[{\citenamefont{Chetyrkin and Grozin}(2003)}]{Chetyrkin:2003vi}
\bibinfo{author}{\bibfnamefont{K.~G.} \bibnamefont{Chetyrkin}}
  \bibnamefont{and} \bibinfo{author}{\bibfnamefont{A.~G.}
  \bibnamefont{Grozin}}, \bibinfo{journal}{Nucl. Phys. B}
  \textbf{\bibinfo{volume}{666}}, \bibinfo{pages}{289} (\bibinfo{year}{2003}),
  \eprint{hep-ph/0303113}.

\bibitem[{\citenamefont{Braun et~al.}(2020)\citenamefont{Braun, Chetyrkin, and
  Kniehl}}]{Braun:2020ymy}
\bibinfo{author}{\bibfnamefont{V.~M.} \bibnamefont{Braun}},
  \bibinfo{author}{\bibfnamefont{K.~G.} \bibnamefont{Chetyrkin}},
  \bibnamefont{and} \bibinfo{author}{\bibfnamefont{B.~A.}
  \bibnamefont{Kniehl}}, \bibinfo{journal}{JHEP} \textbf{\bibinfo{volume}{07}},
  \bibinfo{pages}{161} (\bibinfo{year}{2020}), \eprint{2004.01043}.

\bibitem[{\citenamefont{Grozin}(2024)}]{Grozin:2023dlk}
\bibinfo{author}{\bibfnamefont{A.}~\bibnamefont{Grozin}},
  \bibinfo{journal}{JHEP} \textbf{\bibinfo{volume}{02}}, \bibinfo{pages}{198}
  (\bibinfo{year}{2024}), \eprint{2311.09894}.

\bibitem[{\citenamefont{Collins and Soper}(1981)}]{Collins:1981uk}
\bibinfo{author}{\bibfnamefont{J.~C.} \bibnamefont{Collins}} \bibnamefont{and}
  \bibinfo{author}{\bibfnamefont{D.~E.} \bibnamefont{Soper}},
  \bibinfo{journal}{Nucl. Phys. B} \textbf{\bibinfo{volume}{193}},
  \bibinfo{pages}{381} (\bibinfo{year}{1981}), \bibinfo{note}{[Erratum:
  Nucl.Phys.B 213, 545 (1983)]}.

\bibitem[{\citenamefont{Ji et~al.}(2005)\citenamefont{Ji, Ma, and
  Yuan}}]{Ji:2004wu}
\bibinfo{author}{\bibfnamefont{X.-d.} \bibnamefont{Ji}},
  \bibinfo{author}{\bibfnamefont{J.-p.} \bibnamefont{Ma}}, \bibnamefont{and}
  \bibinfo{author}{\bibfnamefont{F.}~\bibnamefont{Yuan}},
  \bibinfo{journal}{Phys. Rev. D} \textbf{\bibinfo{volume}{71}},
  \bibinfo{pages}{034005} (\bibinfo{year}{2005}), \eprint{hep-ph/0404183}.

\bibitem[{\citenamefont{Korchemskaya and
  Korchemsky}(1992)}]{Korchemskaya:1992je}
\bibinfo{author}{\bibfnamefont{I.~A.} \bibnamefont{Korchemskaya}}
  \bibnamefont{and} \bibinfo{author}{\bibfnamefont{G.~P.}
  \bibnamefont{Korchemsky}}, \bibinfo{journal}{Phys. Lett. B}
  \textbf{\bibinfo{volume}{287}}, \bibinfo{pages}{169} (\bibinfo{year}{1992}).

\bibitem[{\citenamefont{Moch et~al.}(2004)\citenamefont{Moch, Vermaseren, and
  Vogt}}]{Moch:2004pa}
\bibinfo{author}{\bibfnamefont{S.}~\bibnamefont{Moch}},
  \bibinfo{author}{\bibfnamefont{J.~A.~M.} \bibnamefont{Vermaseren}},
  \bibnamefont{and} \bibinfo{author}{\bibfnamefont{A.}~\bibnamefont{Vogt}},
  \bibinfo{journal}{Nucl. Phys. B} \textbf{\bibinfo{volume}{688}},
  \bibinfo{pages}{101} (\bibinfo{year}{2004}), \eprint{hep-ph/0403192}.

\bibitem[{\citenamefont{Henn et~al.}(2020)\citenamefont{Henn, Korchemsky, and
  Mistlberger}}]{Henn:2019swt}
\bibinfo{author}{\bibfnamefont{J.~M.} \bibnamefont{Henn}},
  \bibinfo{author}{\bibfnamefont{G.~P.} \bibnamefont{Korchemsky}},
  \bibnamefont{and}
  \bibinfo{author}{\bibfnamefont{B.}~\bibnamefont{Mistlberger}},
  \bibinfo{journal}{JHEP} \textbf{\bibinfo{volume}{04}}, \bibinfo{pages}{018}
  (\bibinfo{year}{2020}), \eprint{1911.10174}.

\bibitem[{\citenamefont{von Manteuffel et~al.}(2020)\citenamefont{von
  Manteuffel, Panzer, and Schabinger}}]{vonManteuffel:2020vjv}
\bibinfo{author}{\bibfnamefont{A.}~\bibnamefont{von Manteuffel}},
  \bibinfo{author}{\bibfnamefont{E.}~\bibnamefont{Panzer}}, \bibnamefont{and}
  \bibinfo{author}{\bibfnamefont{R.~M.} \bibnamefont{Schabinger}},
  \bibinfo{journal}{Phys. Rev. Lett.} \textbf{\bibinfo{volume}{124}},
  \bibinfo{pages}{162001} (\bibinfo{year}{2020}), \eprint{2002.04617}.

\bibitem[{\citenamefont{Grozin}(2023)}]{Grozin:2022umo}
\bibinfo{author}{\bibfnamefont{A.}~\bibnamefont{Grozin}},
  \bibinfo{journal}{Int. J. Mod. Phys.} \textbf{\bibinfo{volume}{38}}
  (\bibinfo{year}{2023}), \eprint{2212.05290}.

\bibitem[{\citenamefont{Ji et~al.}(2024)\citenamefont{Ji, Liu, Su, and
  Zhang}}]{Ji:2024hit}
\bibinfo{author}{\bibfnamefont{X.}~\bibnamefont{Ji}},
  \bibinfo{author}{\bibfnamefont{Y.}~\bibnamefont{Liu}},
  \bibinfo{author}{\bibfnamefont{Y.}~\bibnamefont{Su}}, \bibnamefont{and}
  \bibinfo{author}{\bibfnamefont{R.}~\bibnamefont{Zhang}}
  (\bibinfo{year}{2024}), \eprint{2410.12910}.

\bibitem[{\citenamefont{Becher et~al.}(2007)\citenamefont{Becher, Neubert, and
  Pecjak}}]{Becher:2006mr}
\bibinfo{author}{\bibfnamefont{T.}~\bibnamefont{Becher}},
  \bibinfo{author}{\bibfnamefont{M.}~\bibnamefont{Neubert}}, \bibnamefont{and}
  \bibinfo{author}{\bibfnamefont{B.~D.} \bibnamefont{Pecjak}},
  \bibinfo{journal}{JHEP} \textbf{\bibinfo{volume}{01}}, \bibinfo{pages}{076}
  (\bibinfo{year}{2007}), \eprint{hep-ph/0607228}.

\bibitem[{\citenamefont{Br\"user et~al.}(2020)\citenamefont{Br\"user, Liu, and
  Stahlhofen}}]{Bruser:2019yjk}
\bibinfo{author}{\bibfnamefont{R.}~\bibnamefont{Br\"user}},
  \bibinfo{author}{\bibfnamefont{Z.~L.} \bibnamefont{Liu}}, \bibnamefont{and}
  \bibinfo{author}{\bibfnamefont{M.}~\bibnamefont{Stahlhofen}},
  \bibinfo{journal}{JHEP} \textbf{\bibinfo{volume}{03}}, \bibinfo{pages}{071}
  (\bibinfo{year}{2020}), \eprint{1911.04494}.

\bibitem[{\citenamefont{Moult et~al.}(2022)\citenamefont{Moult, Zhu, and
  Zhu}}]{Moult:2022xzt}
\bibinfo{author}{\bibfnamefont{I.}~\bibnamefont{Moult}},
  \bibinfo{author}{\bibfnamefont{H.~X.} \bibnamefont{Zhu}}, \bibnamefont{and}
  \bibinfo{author}{\bibfnamefont{Y.~J.} \bibnamefont{Zhu}},
  \bibinfo{journal}{JHEP} \textbf{\bibinfo{volume}{08}}, \bibinfo{pages}{280}
  (\bibinfo{year}{2022}), \eprint{2205.02249}.

\bibitem[{\citenamefont{Bruno et~al.}(2015)}]{Bruno:2014jqa}
\bibinfo{author}{\bibfnamefont{M.}~\bibnamefont{Bruno}} \bibnamefont{et~al.},
  \bibinfo{journal}{JHEP} \textbf{\bibinfo{volume}{02}}, \bibinfo{pages}{043}
  (\bibinfo{year}{2015}), \eprint{1411.3982}.

\bibitem[{\citenamefont{Bali et~al.}(2023)\citenamefont{Bali, Collins,
  Heybrock, L\"offler, R\"odl, S\"oldner, and Weish\"aupl}}]{Bali:2023sdi}
\bibinfo{author}{\bibfnamefont{G.~S.} \bibnamefont{Bali}},
  \bibinfo{author}{\bibfnamefont{S.}~\bibnamefont{Collins}},
  \bibinfo{author}{\bibfnamefont{S.}~\bibnamefont{Heybrock}},
  \bibinfo{author}{\bibfnamefont{M.}~\bibnamefont{L\"offler}},
  \bibinfo{author}{\bibfnamefont{R.}~\bibnamefont{R\"odl}},
  \bibinfo{author}{\bibfnamefont{W.}~\bibnamefont{S\"oldner}},
  \bibnamefont{and}
  \bibinfo{author}{\bibfnamefont{S.}~\bibnamefont{Weish\"aupl}}
  (\bibinfo{collaboration}{RQCD}), \bibinfo{journal}{Phys. Rev. D}
  \textbf{\bibinfo{volume}{108}}, \bibinfo{pages}{034512}
  (\bibinfo{year}{2023}), \eprint{2305.04717}.

\bibitem[{\citenamefont{Martinelli and Sachrajda}(1989)}]{Martinelli:1988rr}
\bibinfo{author}{\bibfnamefont{G.}~\bibnamefont{Martinelli}} \bibnamefont{and}
  \bibinfo{author}{\bibfnamefont{C.~T.} \bibnamefont{Sachrajda}},
  \bibinfo{journal}{Nucl. Phys. B} \textbf{\bibinfo{volume}{316}},
  \bibinfo{pages}{355} (\bibinfo{year}{1989}).

\bibitem[{\citenamefont{Bhattacharya et~al.}(2014)\citenamefont{Bhattacharya,
  Cohen, Gupta, Joseph, Lin, and Yoon}}]{Bhattacharya:2013ehc}
\bibinfo{author}{\bibfnamefont{T.}~\bibnamefont{Bhattacharya}},
  \bibinfo{author}{\bibfnamefont{S.~D.} \bibnamefont{Cohen}},
  \bibinfo{author}{\bibfnamefont{R.}~\bibnamefont{Gupta}},
  \bibinfo{author}{\bibfnamefont{A.}~\bibnamefont{Joseph}},
  \bibinfo{author}{\bibfnamefont{H.-W.} \bibnamefont{Lin}}, \bibnamefont{and}
  \bibinfo{author}{\bibfnamefont{B.}~\bibnamefont{Yoon}},
  \bibinfo{journal}{Phys. Rev.} \textbf{\bibinfo{volume}{D89}},
  \bibinfo{pages}{094502} (\bibinfo{year}{2014}), \eprint{1306.5435}.

\bibitem[{\citenamefont{Ji et~al.}(2021{\natexlab{d}})\citenamefont{Ji, Liu,
  Sch\"afer, Wang, Yang, Zhang, and Zhao}}]{Ji:2020brr}
\bibinfo{author}{\bibfnamefont{X.}~\bibnamefont{Ji}},
  \bibinfo{author}{\bibfnamefont{Y.}~\bibnamefont{Liu}},
  \bibinfo{author}{\bibfnamefont{A.}~\bibnamefont{Sch\"afer}},
  \bibinfo{author}{\bibfnamefont{W.}~\bibnamefont{Wang}},
  \bibinfo{author}{\bibfnamefont{Y.-B.} \bibnamefont{Yang}},
  \bibinfo{author}{\bibfnamefont{J.-H.} \bibnamefont{Zhang}}, \bibnamefont{and}
  \bibinfo{author}{\bibfnamefont{Y.}~\bibnamefont{Zhao}},
  \bibinfo{journal}{Nucl. Phys. B} \textbf{\bibinfo{volume}{964}},
  \bibinfo{pages}{115311} (\bibinfo{year}{2021}{\natexlab{d}}),
  \eprint{2008.03886}.

\bibitem[{\citenamefont{Delcarro et~al.}(2018)\citenamefont{Delcarro,
  Bacchetta, Pisano, Radici, and Signori}}]{Delcarro:2018lbr}
\bibinfo{author}{\bibfnamefont{F.}~\bibnamefont{Delcarro}},
  \bibinfo{author}{\bibfnamefont{A.}~\bibnamefont{Bacchetta}},
  \bibinfo{author}{\bibfnamefont{C.}~\bibnamefont{Pisano}},
  \bibinfo{author}{\bibfnamefont{M.}~\bibnamefont{Radici}}, \bibnamefont{and}
  \bibinfo{author}{\bibfnamefont{A.}~\bibnamefont{Signori}},
  \bibinfo{journal}{PoS} \textbf{\bibinfo{volume}{DIS2018}},
  \bibinfo{pages}{219} (\bibinfo{year}{2018}).

\bibitem[{\citenamefont{Babich et~al.}(2010)\citenamefont{Babich, Brannick,
  Brower, Clark, Manteuffel, McCormick, Osborn, and Rebbi}}]{Babich:2010qb}
\bibinfo{author}{\bibfnamefont{R.}~\bibnamefont{Babich}},
  \bibinfo{author}{\bibfnamefont{J.}~\bibnamefont{Brannick}},
  \bibinfo{author}{\bibfnamefont{R.~C.} \bibnamefont{Brower}},
  \bibinfo{author}{\bibfnamefont{M.~A.} \bibnamefont{Clark}},
  \bibinfo{author}{\bibfnamefont{T.~A.} \bibnamefont{Manteuffel}},
  \bibinfo{author}{\bibfnamefont{S.~F.} \bibnamefont{McCormick}},
  \bibinfo{author}{\bibfnamefont{J.~C.} \bibnamefont{Osborn}},
  \bibnamefont{and} \bibinfo{author}{\bibfnamefont{C.}~\bibnamefont{Rebbi}},
  \bibinfo{journal}{Phys. Rev. Lett.} \textbf{\bibinfo{volume}{105}},
  \bibinfo{pages}{201602} (\bibinfo{year}{2010}), \eprint{1005.3043}.

\bibitem[{\citenamefont{Osborn et~al.}(2010)\citenamefont{Osborn, Babich,
  Brannick, Brower, Clark, Cohen, and Rebbi}}]{Osborn:2010mb}
\bibinfo{author}{\bibfnamefont{J.~C.} \bibnamefont{Osborn}},
  \bibinfo{author}{\bibfnamefont{R.}~\bibnamefont{Babich}},
  \bibinfo{author}{\bibfnamefont{J.}~\bibnamefont{Brannick}},
  \bibinfo{author}{\bibfnamefont{R.~C.} \bibnamefont{Brower}},
  \bibinfo{author}{\bibfnamefont{M.~A.} \bibnamefont{Clark}},
  \bibinfo{author}{\bibfnamefont{S.~D.} \bibnamefont{Cohen}}, \bibnamefont{and}
  \bibinfo{author}{\bibfnamefont{C.}~\bibnamefont{Rebbi}},
  \bibinfo{journal}{PoS} \textbf{\bibinfo{volume}{LATTICE2010}},
  \bibinfo{pages}{037} (\bibinfo{year}{2010}), \eprint{1011.2775}.

\bibitem[{\citenamefont{Edwards and Joo}(2005)}]{Edwards:2004sx}
\bibinfo{author}{\bibfnamefont{R.~G.} \bibnamefont{Edwards}} \bibnamefont{and}
  \bibinfo{author}{\bibfnamefont{B.}~\bibnamefont{Joo}}
  (\bibinfo{collaboration}{SciDAC, LHPC, UKQCD}), \bibinfo{journal}{Nucl. Phys.
  Proc. Suppl.} \textbf{\bibinfo{volume}{140}}, \bibinfo{pages}{832}
  (\bibinfo{year}{2005}), \bibinfo{note}{[,832(2004)]},
  \eprint{hep-lat/0409003}.

\end{thebibliography}

\clearpage

\end{document}